\documentclass[pra,twocolumn,amsmath,amssymb,superscriptaddress]{revtex4-1}
\usepackage{epsfig,amsmath}
\usepackage{subfigure}
\usepackage{graphicx}
\usepackage{dcolumn}
\usepackage{stmaryrd}
\usepackage{mathrsfs}
\usepackage{pifont}
\usepackage{amsthm}
\usepackage{amssymb}
\usepackage{bm}
\usepackage{latexsym}
\usepackage[colorlinks=true,linkcolor=blue,citecolor=blue]{hyperref}
\usepackage{color}
\usepackage{epstopdf}
\usepackage{booktabs}
\usepackage{subfigure}
\usepackage{float}
\usepackage{booktabs,multirow}

\begin{document}

\title{Realizing a mechanical dynamical Casimir effect with a low-frequency oscillator}

\author{Tian-hao Jiang}
\affiliation{School of Physics, Zhejiang University, Hangzhou 310027, Zhejiang, China}

\author{Jun Jing}
\email{Email address: jingjun@zju.edu.cn}
\affiliation{School of Physics, Zhejiang University, Hangzhou 310027, Zhejiang, China}
	
\date{\today}

\begin{abstract}
We propose to realize a mechanical dynamical Casimir effect (MDCE) in a hybrid optomechanical system consisting of a cavity mode, a low-frequency mechanical oscillator, and a two-level atomic system. Described by the effective Hamiltonian, the mechanical energy is directly converted to the photons through a three-wave-mixing mechanism. It is not a quantum simulation of a parametric DCE such as in superconducting circuits. Using a master-equation approach, we analyze the system dynamics in various regimes with respect to the ratio of the effective coupling strength and the loss rate of the system. The dynamics under the strong-coupling regime confirms various three-wave-mixing processes for creating photons by annihilation of mechanical and atomic excitations. Under the weak-coupling regime, a continuous production of photons can be demonstrated by driving both the mechanical oscillator and atom. By virtue of an atom of tunable frequency, our method avoids using the high-frequency mechanical oscillator, which is required for the conventional DCE in optomechanical systems under the double-photon resonance yet is out of reach of experiment. It is found that the mechanical frequency can be about two orders smaller than the cavity mode in yielding a remarkable flux of DCE photons.
\end{abstract}

\maketitle

\section{Introduction}\label{Introduction}

The uncertainty principle in quantum mechanics leads to the vacuum state filled with constant activity. Fluctuations in the quantum vacuum give rise to a myriad of particles that appear to materialize and vanish in an instant~\cite{hawking1974Black,hawking1975Particle,unruh1976Notes,
wilson2011Observation,nation2012Colloquium,benenti2014Dynamical,macri2018Nonperturbative,distefano2019Interaction}. Distinguished examples include Hawking radiation~\cite{hawking1974Black,hawking1975Particle}, the Unruh effect~\cite{unruh1976Notes}, the static Casimir effect~\cite{casimir1948Attraction,sparnaay1958Measurements,lamoreaux1997Demonstration}, and the dynamical Casimir effect (DCE)~\cite{wilson2011Observation,nation2012Colloquium,benenti2014Dynamical,macri2018Nonperturbative,
distefano2019Interaction,lahteenmaki2013Dynamical}.

The static Casimir effect, first proposed by Casimir in~\cite{casimir1948Attraction}, predicted an attractive force between two perfect and electrically neutral metal plates. This attraction arises due to quantum fluctuations in the vacuum and the zero-point energy filtered out by the boundary condition. In~\cite{sparnaay1958Measurements} Sparnaay reported the first experimental attempt to measure this effect. His result confirmed the attractive force between the metal plates. Lamoreaux conducted a more precise measurement using a torsion pendulum~\cite{lamoreaux1997Demonstration}, whose results differed from the Casimir's theoretical prediction by no more than $5\%$ . Distinct from the static Casimir effect, a time-varying dynamical Casimir effect was proposed~\cite{moore2003Quantum}, which predicted that a cavity with a moving mirror can create photons when the vibration frequency of the mirror is twice of the cavity mode~\cite{lambrecht1996Motion}. In general, the DCE can be understood as a macroscopic phenomenon resulting from nonadiabatic changes in the boundary positions or properties of quantum fields~\cite{dodonov2020Fifty}, regardless of multiple modes or a single mode. For example, the DCE has been discussed in optomechanical systems of a single cavity mode~\cite{macri2018Nonperturbative,qin2019Emission}, which describe the interaction between the cavity field and mechanical oscillator from a fully-quantum-mechanical perspective.

Explorations of the DCE have been extended from the cosmological problems, such as particle creation due to boundary motion that constrains the field in the curved space-time or in the presence of gravitational fields~\cite{brevik2000Dynamical,ruser2007Dynamical,lock2017Dynamical,ottewill1988Radiation}, to quantum information science. For example, the DCE can be used to cool the cavity wall in the presence of a nonvanishing temperature gradient between the wall and the cavity~\cite{ferreri2024Phononphoton}. A quantum Otto heat engine has been implemented with the DCE~\cite{ferreri2023Quantum}, which operates within a cavity consisting of two oscillating mirrors confining an optical field. The mutual transformation between phonons and photons by the DCE leads to an energy exchange between a cold bath and a hot bath, in which the walls and the single field mode constitute the working substance of the engine. Other applications of the DCE include entanglement generation~\cite{felicetti2014Dynamical,busch2014Quantum,aron2014Steadystate,
rossatto2016Entangling,agusti2019Entanglement}, quantum gate construction~\cite{friis2012Quantum,bruschi2013Relativistic}, quantum steering~\cite{sabin2015Generation}, and quantum communication~\cite{benenti2014Dynamical}( see,e.g., Refs.~\cite{benenti2014Dynamical,felicetti2014Dynamical,aron2014Steadystate,rossatto2016Entangling,agusti2019Entanglement} , which discuss the single field mode).

Direct realization and detection of the DCE challenge the current experiments in optomechanical systems~\cite{qin2019Emission}, primarily due to the difficulty in moving the massive mirrors with a sufficiently high frequency. The double-photon resonance condition requires that the mechanical frequency $\omega_m$ is about twice the cavity-mode frequency $\omega_c$~\cite{lambrecht1996Motion}. To produce a typical microwave photon of $\omega_c/2\pi\sim 5$ GHz through the DCE, the frequency of the mechanical oscillator should be about $\omega_m/2\pi\sim 10$ GHz, which seems beyond the experimental reach~\cite{oconnell2010Quantum}. A relevant challenge is the resonant driving used to constantly excite the phonon mode of high frequency in optomechanical systems~\cite{macri2018Nonperturbative}. The problem of realization has been addressed by the proposals for the parametric dynamical Casimir effect (PDCE)~\cite{johansson2009Dynamical,johansson2010Dynamical,crocce2004Model,nation2012Colloquium}, which suggest the simulation of the boundary conditions by effective motion. For example, microwave photons can be emitted from vacuum by fast modulation over the flux through a superconducting quantum-interference device~\cite{johansson2009Dynamical,johansson2010Dynamical,dodonov2010Current,wilson2011Observation} or over the refraction index of a Josephson metamaterial~\cite{lahteenmaki2013Dynamical}. The problem of detection can be solved by a nearly-quantum-limited photodetection scheme based on superradiant amplification~\cite{kim2006Detectability,brownell2008Modelling}. In particular, Casimir photons are detected through their interaction with ultracold alkali-metal atoms prepared in an inverted population of hyperfine states.

With respect to a direct conversion of mechanical energy into the photon output rather than a simulation through the PDCE, several methods are designed to realize the mechanical dynamical Casimir effect (MDCE)~\cite{sassaroli1994Photon,haro2006Hamiltonian,kim2006Detectability,brownell2008Modelling,macri2018Nonperturbative,
settineri2019Conversion,qin2019Emission,lan2024Dynamical}. An amplified MDCE can be observed in the squeezed frame~\cite{qin2019Emission}. It is realized by a detuned driving and a resonant coupling between the mechanical oscillator and the squeezed cavity mode. Photons can be continuously generated by coupling the mechanical oscillator to a phononic reservoir with a finite temperature~\cite{settineri2019Conversion}. The standard condition of the double-photon resonance can be relaxed to $k\omega_m=2\omega_c$ (with $k$ an integer) by using an ultrastrong optomechanical coupling $g/\omega_c\geq 0.1$~\cite{macri2018Nonperturbative}. Despite this relaxation indicating a significant advancement, $k$ is severely limited. A substantial reduction in the mechanical frequency thus remains desired to generate DCE photons.

In this work we propose the realization of the MDCE in a hybrid qubit-photon-phonon system, consisting of a single-mode cavity with a movable mirror and a two-level system~\cite{wang2023Coherent}. Using the effective Hamiltonian theory~\cite{garziano2016One,macri2018Simple,qi2020Generating}, we find that the three-wave-mixing process constitutes the underlying mechanism of the MDCE, which accounts for the annihilation of the phonon and atomic excitations and the creation of microwave photons in the leaky cavity. The dynamics shows that the MDCE can be realized in a range of mechanical frequency much lower than that in existing works, which also solves the problem of high-frequency driving. As the mechanical frequency decreases, our calculation of a signal-to-noise ratio allows for discrimination in the triggering sources for photons: either the MDCE or the Rabi oscillation between the cavity mode and the qubit under driving. The role played by the MDCE is confirmed to be dominant even when the phonon frequency is as low as a small percentage of the photon frequency.

The rest of this paper is structured as follows. In Sec.~\ref{description of the model} we describe the hybrid system and the full Hamiltonian. We obtain the parametric conditions for the three-wave-mixing process, which generates a photon in the cavity and simultaneously annihilates atomic excitation and a phonon. The relevant effective Hamiltonian is justified by comparing the analytical results and the numerical simulation. We discuss the population dynamics of photons, phonons, and qubit in the strong-coupling and weak-coupling regimes in Secs.~\ref{Strong-coupling regime} and \ref{Weak-coupling regime}, respectively. In Sec.~\ref{Strong-coupling regime} it is shown that the photon output is triggered by the three-wave-mixing mechanism. In Sec.~\ref{Weak-coupling regime} a continuous output of photons is realized through continuous driving on the mechanical motion and qubit. We summarize the paper in Sec.~\ref{conclusion}.

\section{Theoretical model}\label{description of the model}

\begin{figure}[htbp]
\centering
\includegraphics[width=0.95\linewidth]{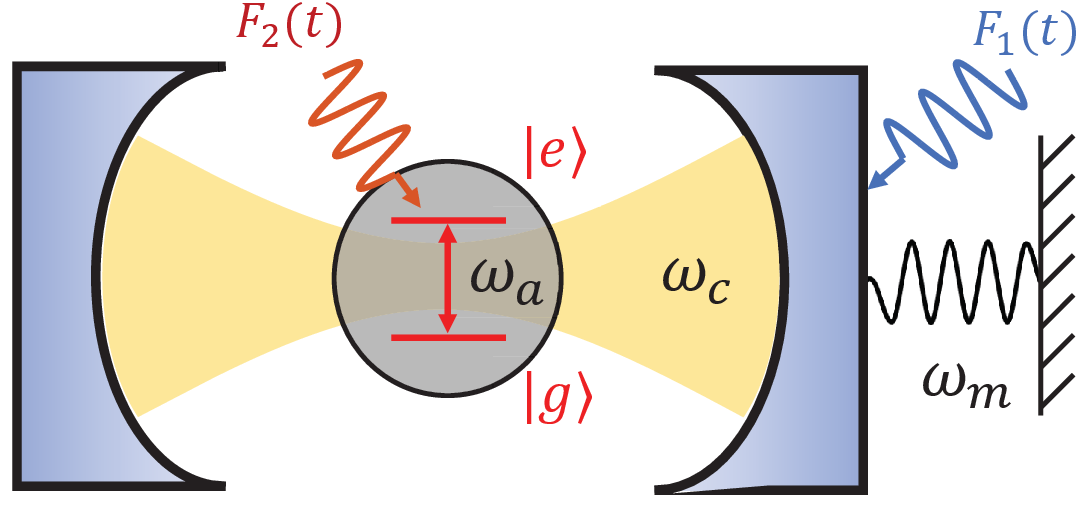}
\caption{Diagram of a hybrid system consisting of a qubit and a single-mode cavity with a movable mirror. Their frequencies are tuned to satisfy the three-wave mixing condition: $\omega_a+\omega_m\approx\omega_c$. Here $F_1(t)$ and $F_2(t)$ are the Rabi frequencies for the external driving on the mechanical oscillator and the qubit, respectively.}\label{model}
\end{figure}

Consider a cavity optomechanical setup composed of three components: a cavity optical resonator, a mechanical oscillator, and a two-level atomic system (see Fig.~\ref{model}). The full Hamiltonian consists of the free Hamiltonian and the interaction Hamiltonian.
\begin{equation}\label{fullH}
  H=H_0+V.
\end{equation}
The free Hamiltonian ($\hbar=1$) reads
\begin{equation}
H_0=\omega_a\sigma_+\sigma_-+\omega_c a^\dagger a+\omega_m b^\dagger b,
\end{equation}
where the level splitting of the atom is $\omega_a$ ; the eigenfrequencies for the photon and phonon are $\omega_c$ and $\omega_m$, respectively; $\sigma_+$ and $\sigma_-$ are the raising and lowering operators for the qubit, respectively; and $a$ ($b$) and $a^\dagger$ ($b^\dagger$) are annihilation and creation operators for the photon (phonon), respectively. The interaction Hamiltonian~\cite{wang2023Coherent} $V=V_{\rm AF}+V_{\rm OM}+V_{\rm DCE}$ consists of the coupling between the atom and cavity mode $V_{\rm AF}$ and the coupling between the cavity mode and mechanical mode $V_{\rm OM}+V_{\rm DCE}$, where $V_{\rm OM}$ and $V_{\rm DCE}$ represent the photon-pressure part and the dynamical-Casimir-effect part, respectively. In particular, we have
\begin{equation}\label{interaction}
\begin{aligned}
V_{\rm AF} &=\lambda(a^\dagger + a)(\sigma_+ + \sigma_-),\\
V_{\rm OM}&= g a^\dagger a(b+b^\dagger),\\
V_{\rm DCE}&=\frac{g}{2}(a^2+a^{\dagger 2})(b+b^\dagger),
\end{aligned}
\end{equation}
where $\lambda$ and $g$ denote the atom-cavity and photon-phonon coupling strengths, respectively. Many optomechanical models neglect $V_{\rm DCE}$~\cite{groblacher2009Observation,verhagen2012Quantumcoherent,bochmann2013Nanomechanical} under the approximation where the mechanical frequency is much lower than the cavity frequency~\cite{aspelmeyer2014Cavity}.

The fundamental mechanism in our method for the MDCE is based on the three-wave-mixing process involving the three components of our system, which can be described by an effective Hamiltonian based on the second-order perturbation theory. The eigenstates for the unperturbed Hamiltonian $H_0$ can be expressed by the tensor-product state $|jnm\rangle\equiv |j\rangle\otimes|n\rangle\otimes|m\rangle$, where $|j\rangle$ ($j=g,e$) labels the atom state and $|n\rangle$ ($|m\rangle$) denotes the number state of the cavity (mechanical) mode. The perturbative Hamiltonian $V$ in Eq.~(\ref{interaction}), which is in charge of the photon generation by the deexcitation of atom and phonons, yields the transition between the states $|e,n,m+1\rangle$ and $|g,n+1,m\rangle$. Then the effective Hamiltonian spanned by $\{|e,n,m+1\rangle,|g,n+1,m\rangle\}$ for arbitrary nonnegative integers $n$ and $m$ can be extracted from the full Hamiltonian~(\ref{fullH}). The effective coupling strength or energy shift between any eigenstates $|i\rangle$ and $|f\rangle$ of the unperturbed Hamiltonian $H_0$ can be obtained by
\begin{equation}\label{standard perturbation theory}
\Omega=V_{fi}+\sum_{n\neq i,f}\frac{V_{fn}V_{ni}}{\omega_i-\omega_n},
\end{equation}
where $V_{nm}\equiv\langle n|V|m\rangle$ and $\omega_n$ is the eigenenergy of the eigenstate $|n\rangle$ for $H_0$. When $|i\rangle\neq|f\rangle$, Eq.~(\ref{standard perturbation theory}) quantifies their coupling strength. Otherwise, it gives rise to the energy-level shift or correction $\epsilon$ of $|i\rangle$.

\begin{figure}[htbp]
\centering
\includegraphics[width=0.95\linewidth]{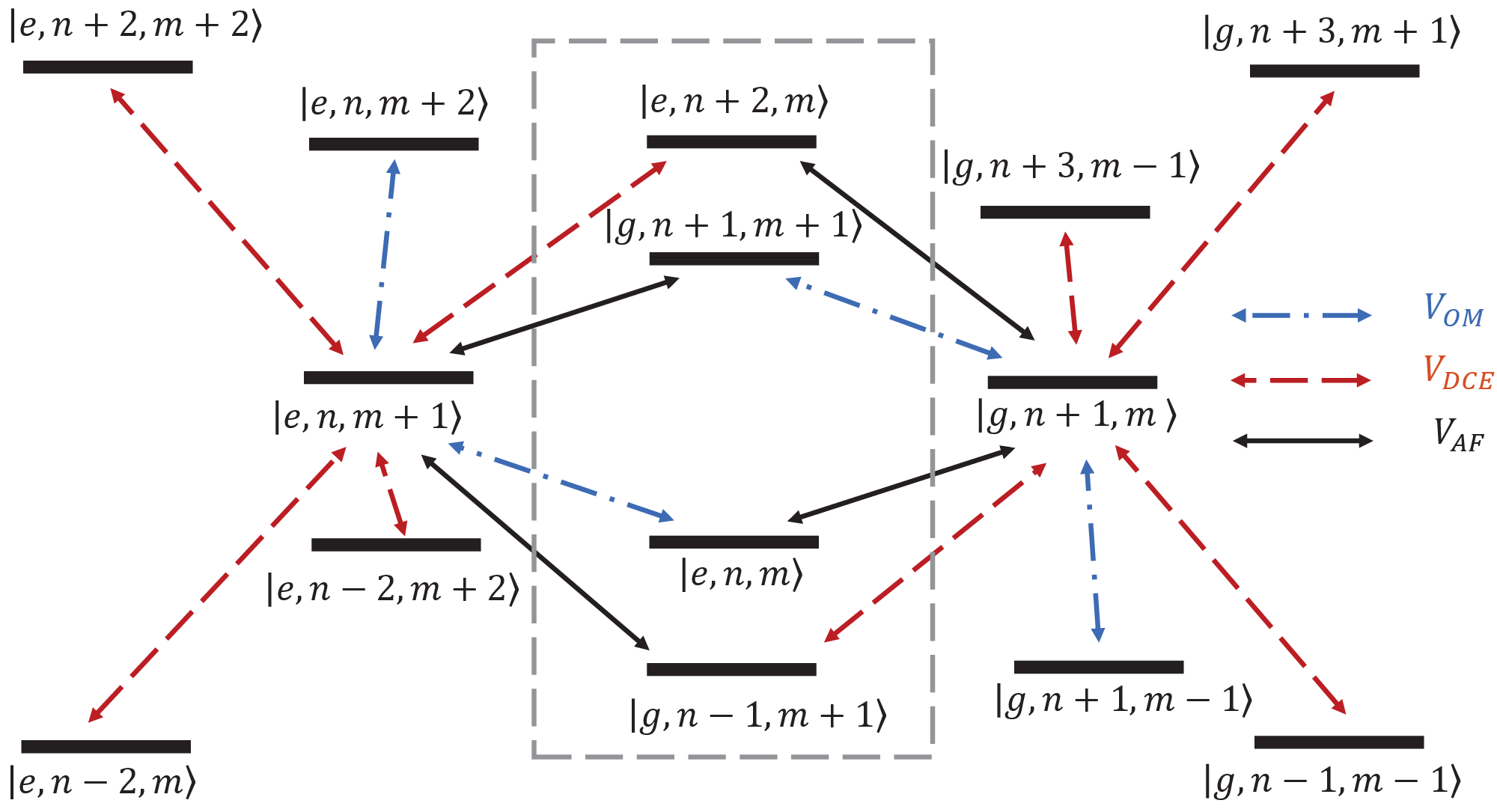}
\caption{All the second-order (two-step) transition paths linking the target eigenstates $|e,n,m+1\rangle$ and $|g,n+1,m\rangle$. The involved levels appear in Eq.~(\ref{standard perturbation theory}) and the transition is determined by the interaction Hamiltonian~(\ref{interaction}). In particular, the blue dot-dashed lines, the red dashed lines, and the black solid lines mark the transitions induced by the standard optomechanical interaction $V_{\rm OM}$, the DCE interaction $V_{\rm DCE}$, and the atom-field interaction $V_{\rm AF}$, respectively.}\label{energy_shift}
\end{figure}

Figure~\ref{energy_shift} serves as a road map for calculating both the effective strength and energy shift, which are exemplified by the second-order paths connecting two typical target eigenstates $|e,n,m+1\rangle$ and $|g,n+1,m\rangle$ and the round-trips for them. According to Eqs.~(\ref{interaction}) and (\ref{standard perturbation theory}), we have four levels (see $|e,n+2,m\rangle$, $|g,n+1,m+1\rangle$, $|e,n,m\rangle$, and $|g,n-1,m+1\rangle$ in the gray dashed box of Fig.~\ref{energy_shift}) to mediate the initial state $|i\rangle=|e,n,m+1\rangle$ and the final state $|f\rangle=|g,n+1,m\rangle$. Thereby, one can obtain the effective coupling strength $g_{\rm eff}$,
\begin{equation}\label{coupling strength}
g_{\rm eff}=-g\lambda\sqrt{n+1}\sqrt{m+1}
\left(\frac{1}{2\omega_c-\omega_m}+\frac{1}{\omega_m}\right).
\end{equation}
It can be verified that $V_{\rm DCE}$ contributes to the first term in Eq.~(\ref{coupling strength}), $V_{\rm OM}$ contributes to the second term, and $V_{\rm AF}$ contributes to both of them.

\begin{figure}[htbp]
\centering
\includegraphics[width=0.95\linewidth]{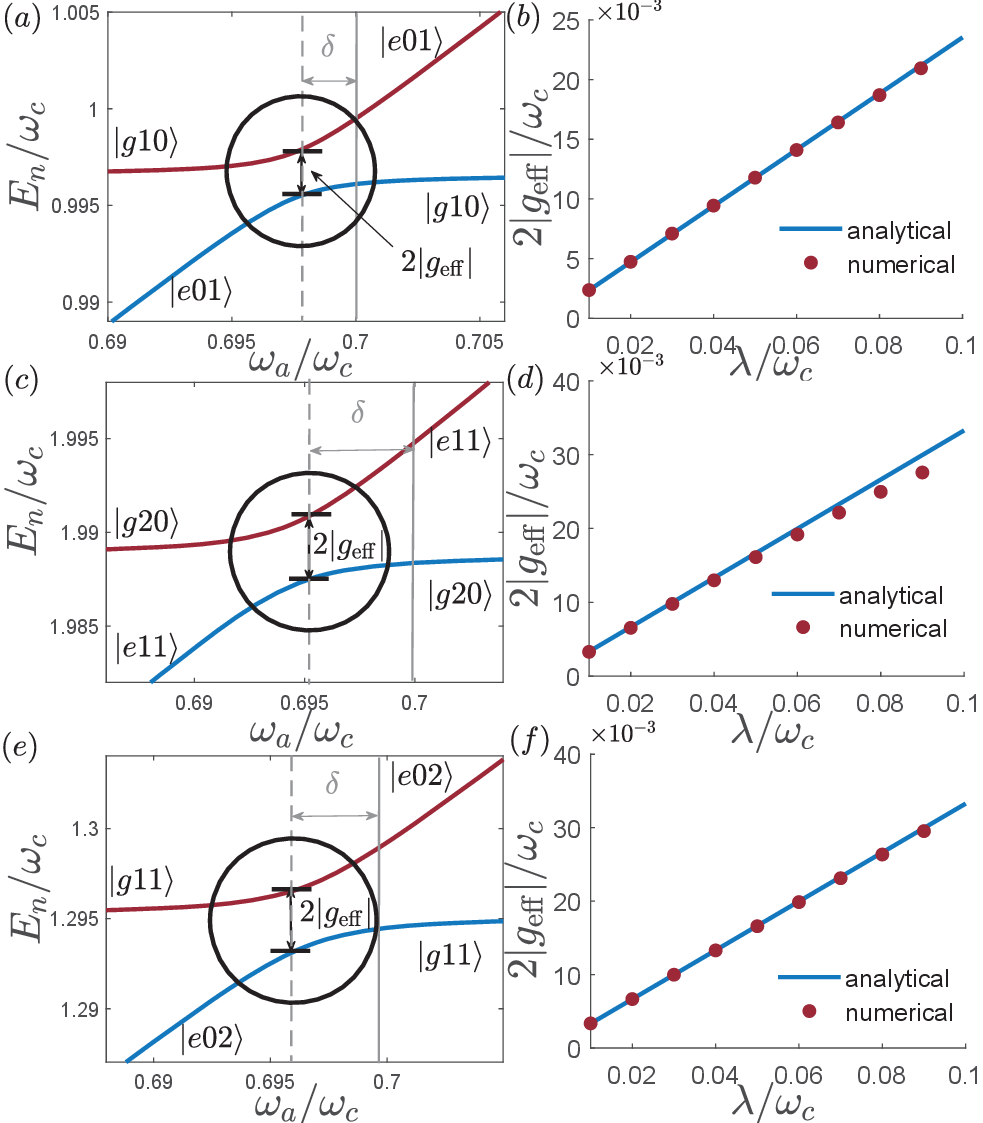}
\caption{(a), (c), and (e) Energy levels and avoided level crossings for various three-wave-mixing processes as a function of the qubit transition frequency $\omega_a$ for $\omega_m=0.3\omega_c$, $g=0.03\omega_c$, and $\lambda=0.01\omega_c$. (b), (d), and (f) Comparison between the numerical evaluation of the normalized effective coupling strengths (red dots) and the corresponding analytical results (blue solid line) for $\omega_m=0.3\omega_c$, $\omega_a=0.7\omega_c$, and $g=0.03\omega_c$.}\label{anticorossing}
\end{figure}

In Fig.~\ref{anticorossing} we demonstrate the avoided level crossings (distinguished by the dark circles) describing transitions between various initial and final states in the diagrams of energy levels $\{E_n\}$ for the full Hamiltonian. In particular, Figs.~\ref{anticorossing}(a),~\ref{anticorossing}(c), and~\ref{anticorossing}(e) plot the target state pairs $|g10\rangle \leftrightarrow |e01\rangle$, $|g20\rangle \leftrightarrow |e11\rangle$, and $|g11\rangle \leftrightarrow |e02\rangle$, respectively. The eigenvalues, which vary with $\omega_a$, are obtained via a standard numerical diagonalization in a truncated Hilbert space of the full Hamiltonian. Around the three-wave-mixing condition, i.e., $\omega_a+\omega_m=\omega_c$, the two states $|e,n,m+1\rangle$ and $|g,n+1,m\rangle$ become nearly degenerate. In addition, the avoided level crossing occurs with a splitting as much as the transition rate $2|g_{\rm eff}|$ for the Rabi oscillation between $|e,n,m+1\rangle$ and $|g,n+1,m\rangle$, which can be obtained by Eq.~(\ref{coupling strength}). When the energy shift $\delta$ is exactly obtained, a perfect Rabi oscillation can be observed, implying that the qubit facilitates the energy conversion from phonon to photon.

Figures~\ref{anticorossing}(b),~\ref{anticorossing}(d), and~\ref{anticorossing}(f) are used to verify the effective coupling strength $g_{\rm eff}$ in Eq.~(\ref{coupling strength}) in terms of the coupling strength $\lambda$ between the qubit and cavity mode. The analytical results are found to match with the numerical ones even when the normalized qubit-photon interaction strength is close to the regime $\lambda/\omega_c\leq 0.1$. Nevertheless, such an ultrastrong coupling is not required by our method in realizing the MDCE.

Similar to the coupling strength $g_{\rm eff}$, the energy shift $\delta$ [distinguished by the gray dashed and solid lines in Figs.~\ref{anticorossing}(a),~\ref{anticorossing}(c), and~\ref{anticorossing}(e)] can be well explained by Fig.~\ref{energy_shift}. For $|g,n+1,m\rangle$, one can find eight round-trips mediated by various levels to correct its energy due to the interaction Hamiltonian~(\ref{interaction}). For example, we have $|g,n+1,m\rangle\rightarrow|g,n+3,m-1\rangle\rightarrow|g,n+1,m\rangle$ from the DCE interaction $V_{\rm DCE}$. Collecting all the relevant contributions with Eq.~(\ref{standard perturbation theory}), one can obtain the energy shift up to the second order of $g$ and $\lambda$,
\begin{equation}\label{epsilon_1}
\begin{aligned}
\epsilon_1=& -\frac{1}{4}g^2\left(\frac{n^2+4nm+5n+6m+6}{2\omega_c+\omega_m}\right. \\
&\left.+\frac{-n^2+4nm+6m-n}{2\omega_c-\omega_m}\right)
-\frac{(n+2)\lambda^2}{\omega_c+\omega_a}\\		&-\frac{(n+1)\lambda^2}{\omega_a-\omega_c}-\frac{(n+1)^2 g^2}{\omega_m}
\end{aligned}
\end{equation}
for $|g,n+1,m\rangle$. Similarly, the energy shift for the state $|e,n,m+1\rangle$ reads,
\begin{equation}\label{epsilon_2}
\begin{aligned}
\epsilon_2=& -\frac{1}{4}g^2\left(\frac{2n^2+4nm+6n+2m+4}{2\omega_c+\omega_m}\right. \\
&\left.+\frac{-n^2+4nm+2m+5n+2}{2\omega_c-\omega_m}\right)
+\frac{n\lambda^2}{\omega_c+\omega_a}\\	&-\frac{(n+1)\lambda^2}{\omega_c-\omega_a}-\frac{n^2 g^2}{\omega_m}.
\end{aligned}
\end{equation}
Equations~(\ref{coupling strength})-(\ref{epsilon_2}) give rise to the effective Hamiltonian in the subspace spanned by $\{|e,n,m+1\rangle,|g,n+1,m\rangle\}$,
\begin{equation}\label{effective Hamiltonian}
\begin{aligned}
H_{\rm eff}=&g_{\rm eff}(|e,n,m+1\rangle\langle g,n+1,m|+{\rm H.c.})\\
&+E_g|g,n+1,m\rangle\langle g,n+1,m|\\
&+E_e|e,n,m+1\rangle\langle e,n,m+1|,
\end{aligned}
\end{equation}
where $E_g=(n+1)\omega_c+m\omega_m+\epsilon_1$ and $E_e=\omega_a+n\omega_c+(m+1)\omega_m+\epsilon_2$ are the eigenenergies of $|g,n+1,m\rangle$ and $|e,n,m+1\rangle$, respectively. To realize a perfect Rabi oscillation between the two target states, the last two terms in Eq.~(\ref{effective Hamiltonian}) should become the identity operator in the corresponding subspace. Thus the resonance condition $E_g=E_e$, i.e., $\omega_c+\epsilon_1=\omega_a+\omega_m+\epsilon_2$, yields the energy shift $\delta$ for the avoided level crossings in Fig.~\ref{anticorossing}:
\begin{equation}\label{frequency shift}
\begin{aligned}
\delta&\equiv\epsilon_2-\epsilon_1=\omega_c-\omega_m-\omega_a \\
&=\frac{g^2}{4} \left(\frac{-n^2-n+4m+2}{2\omega_c+\omega_m}+\frac{-6n+4m-2}{2\omega_c-\omega_m}\right)\\
&+\frac{2(n+1)\lambda^2}{\omega_a-\omega_c} +\frac{(2n+1)g^2}{\omega_m}+\frac{2(n+1)\lambda^2}{\omega_a+\omega_c}.
\end{aligned}
\end{equation}

\begin{figure}[htbp]
\centering
\includegraphics[width=0.95\linewidth]{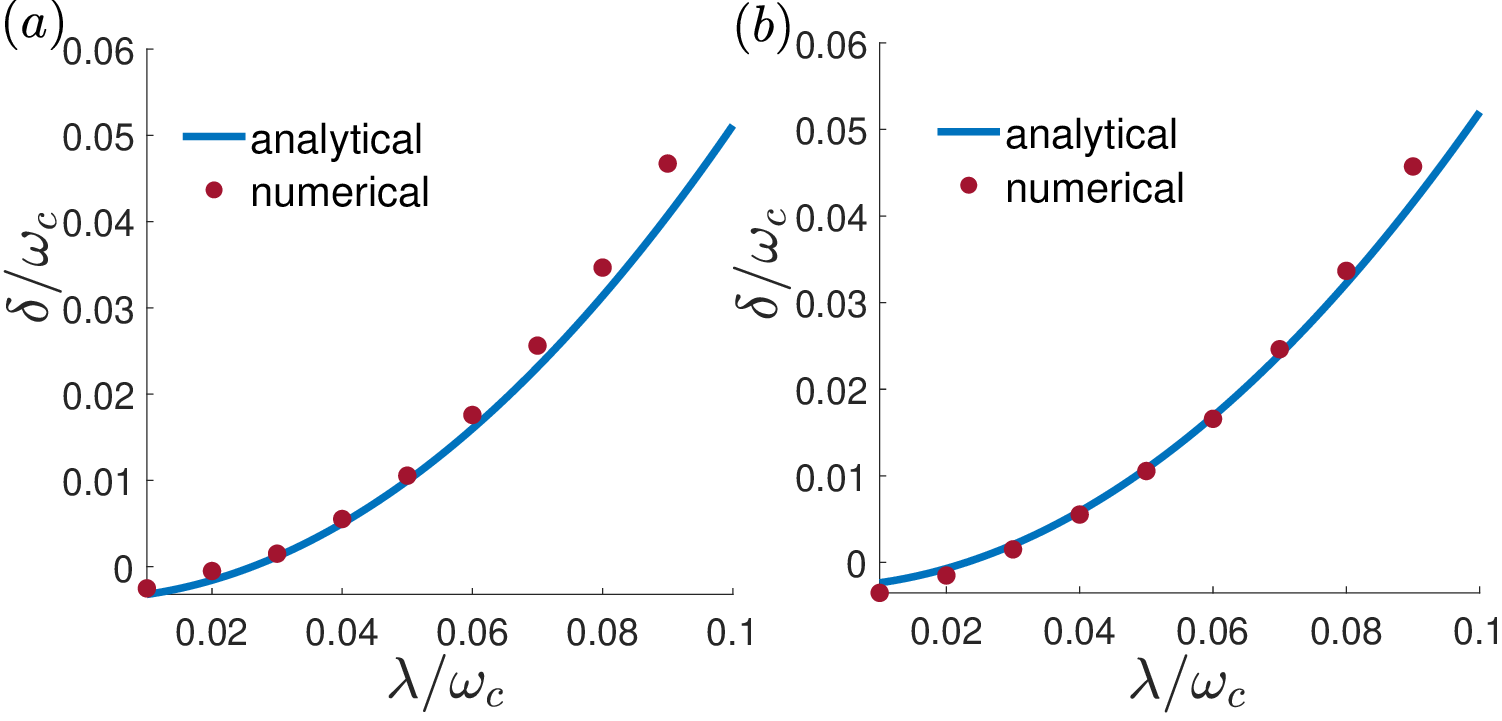}
\caption{Comparison between the numerically calculated normalized energy shift $\delta/\omega_c$ (red dots) and the corresponding analytical results (blue solid line) for the transitions (a) $|g10\rangle\leftrightarrow |e01\rangle$ and (b) $|g11\rangle\leftrightarrow |e02\rangle$. The parameters are the same as in Fig.~\ref{anticorossing}. }\label{delta}
\end{figure}

In Figs.~\ref{delta}(a) and~\ref{delta}(b) we compare the analytical results and the numerical evaluation obtained by Eq.~(\ref{frequency shift}) for the energy shift in the transitions $|g10\rangle\leftrightarrow |e01\rangle$ and $|g11\rangle\leftrightarrow |e02\rangle$, respectively. The calculations are performed under the same conditions as in Fig.~\ref{anticorossing}. It is shown that the analytical results for the energy shift also match the numerical ones when the normalized qubit-photon interaction strength $\lambda/\omega_c\leq 0.1$. In this case, the effective Hamiltonian~(\ref{effective Hamiltonian}) can be expressed as
\begin{equation}\label{H_eff}
H_{\rm eff}\approx g_{\rm eff}(|e,n,m+1\rangle\langle g,n+1,m|+{\rm H.c.}),
\end{equation}
by which the three-wave-mixing process provides the main mechanism for the MDCE emerging in the system dynamics in the following section. It can be witnessed by the Fourier transform about the mean photon number in Sec.~\ref{Strong-coupling regime} and every Fourier peak reveals the transition between a specific pair of target states.

\section{System dynamics}\label{system dynamics}

In this section we explore the system dynamics under the external driving on the mechanical oscillator and the qubit, which is necessary to constantly generate photons in coordination with the three-wave-mixing process supported by the effective Hamiltonian~(\ref{H_eff}). The driving Hamiltonian $H_d(t)$ can be written as
\begin{equation}\label{driving}
H_d(t)=F_1(t)(b^\dagger+b)+F_2(t)(\sigma_+ + \sigma_-),
\end{equation}
where $F_1(t)$ and $F_2(t)$ denote the external force or modulation applied to the movable mirror and the qubit, respectively. Here $F_1(t)$ and $F_2(t)$ are assumed to follow the same formation and are distinct only in the driving frequency.

We employ a Lindblad master equation to evaluate the density-matrix operator $\rho(t)$ of the system under a dissipative environment at zero temperature. It is given by
\begin{equation}\label{master equation}
\begin{aligned}
\dot{\rho}(t)=&i[\rho(t), H+H_d(t)]+\kappa\mathcal{L}[\sigma_-]\rho(t)\\
&+\eta\mathcal{L}[a]\rho(t)+\gamma\mathcal{L}[b]\rho(t),
\end{aligned}
\end{equation}
where $\kappa$, $\eta$, and $\gamma$ are the atomic, photonic, and mechanical loss rates, respectively. In addition, the superoperator $\mathcal{L}$ is defined as
\begin{align}
\mathcal{L}[o]\rho &\equiv \frac{1}{2}\left(2O\rho O^\dagger-O^\dagger O\rho -\rho O^\dagger O\right),\\ O &\equiv \sum_{E_n>E_m}\langle\Psi_m |(o+o^\dagger)|\Psi_n\rangle|\Psi_m\rangle\langle\Psi_n|
\end{align}
where $o=\sigma_-,a,b$ and $|\Psi_n\rangle$ are the eigenvectors of the full Hamiltonian~(\ref{fullH}).

In numerical simulations, the whole system is assumed to be initialized as the ground state $|g00\rangle$. The system parameters follow the setting in Fig.~\ref{anticorossing}(a), i.e., the mechanical frequency $\omega_m=0.3\omega_c$, the atomic splitting $\omega_a\approx0.7\omega_c-\delta$, the qubit-photon coupling strength $\lambda=0.01\omega_c$, and the photon-phonon coupling strength $g=0.03\omega_c$, unless stated otherwise. Substituting them into Eq.~(\ref{coupling strength}), the effective coupling strength is found to be $|g_{\rm eff}|=1.18\times10^{-3}\omega_c$.

In the following, we study the system dynamics in the strong-coupling and weak-coupling regimes when the effective coupling strength $|g_{\rm eff}|$ is higher or lower than the system loss rate, respectively. The former is used to distinguish the mechanism underlying the MDCE. The latter is used to investigate the frequency range of the mechanical oscillator admitted by our method, which is meaningful to avoid the high-frequency requirement in both eigenfrequency and driving frequency for the mechanical oscillator.

In Sec.~\ref{Strong-coupling regime} an ultrafast drive, which is nearly an instant pulse in the temporal envelope, is used to rapidly excite the mechanical oscillator and the atom. Then the system is mainly subject to the time-independent effective Hamiltonian in the subsequent time evolution. Since the three-wave-mixing process is not severely influenced by the comparatively weak dissipation, it can display the mechanically induced vacuum radiation. In contrast, Sec.~\ref{Weak-coupling regime} applies a continuous drive and a comparatively strong dissipation to realize the MDCE with a low-frequency mechanical oscillator. To confirm that the photon output is essentially generated via the MDCE rather than by the direct conversion from the qubit to photon, we compare the stable outputs of the photon number when driving both the mechanical oscillator and qubit and when driving only the qubit. In between the two distinct regimes, the intermediate dynamical behavior is examined by varying the pulse endurance and the loss rate.

\subsection{Strong-coupling regime}\label{Strong-coupling regime}

In the strong-coupling regime, the effective coupling strength overwhelms the system loss [the loss rates in the master equation~(\ref{master equation}) are set as $C=\kappa=\eta=\gamma=10^{-4}\omega_c$] and then one can observe the basic transition determined by the system Hamiltonian during the weak dissipative evolution. The two ultra-fast resonant pulses in Eq.~(\ref{driving}) are set as
\begin{equation}\label{ultrafast}
\begin{aligned}
  F_1(t)&=AG(t-t_0)\cos(\omega_m t), \\
  F_2(t)&=AG(t-t_0)\cos(\omega_a t),
\end{aligned}
\end{equation}
respectively, where $A$ is a dimensionless pulse amplitude or intensity. Here we use a compact Gaussian function as the envelope function of the pulse
\begin{equation}\label{gauss}
  G(t-t_0)=\frac{1}{\sqrt{2\pi}\sigma}\exp\left[-\frac{(t-t_0)^2}{2\sigma^2}\right],
\end{equation}
where $\sigma$ is the standard deviation representing the practical time extension of the pulse and $t_0$ is the central time point. They are set as $\sigma=(16|g_{\rm eff}|)^{-1}$ and $t_0=100/\omega_c$, respectively, unless stated otherwise.

\begin{figure}[htbp]
\centering
\includegraphics[width=0.95\linewidth]{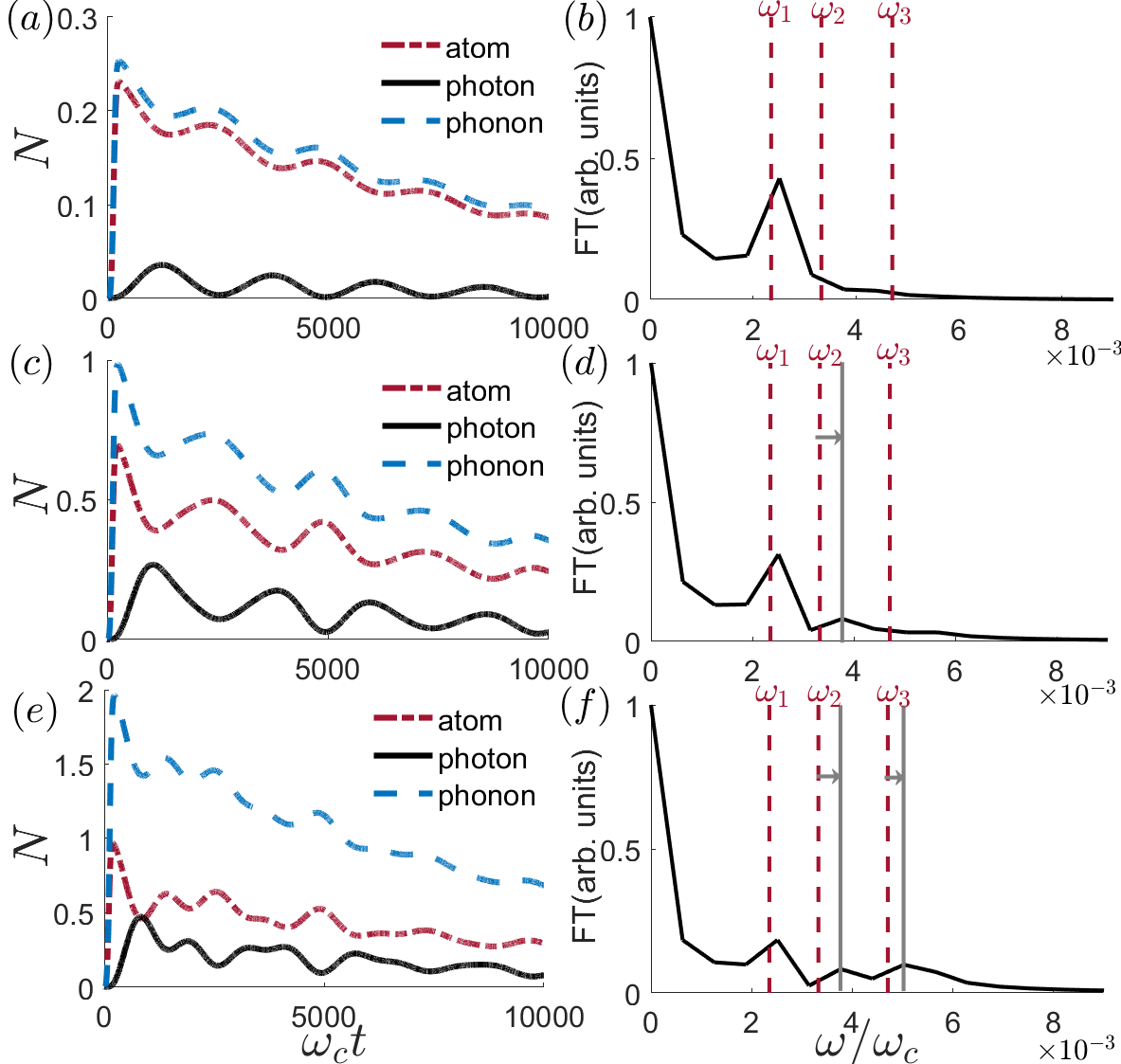}
\caption{Time evolution of the average population of the atom (red dot-dashed line), the mechanical oscillator (blue dashed line), and the cavity mode (black solid line) under the ultrafast pulse in Eq.~(\ref{ultrafast}) with various intensities: (a) $A=\pi/3$, (c) $A=2\pi/3$, and (e) $A=\pi$. (b), (d), and (f) Fourier transformation over the average photon number in (a), (c), and (e), respectively. }\label{dynamic}
\end{figure}

Figure~\ref{dynamic} shows the evolution of the average excitation numbers of three components and the Fourier transformation of the mean photon number under different pulse amplitudes. The Fourier transformation can be defined as
\begin{equation}
F(\omega)=\sum_{k=0}^{M-1} N(k\Delta t) e^{-i(2\pi /M)\omega k \Delta t},
\end{equation}
where $N$ is the mean photon number, $\Delta t$ is the sampling time interval, and $M$ is the sampling number of the mean photon number during the evolution.

In Fig.~\ref{dynamic}(a) the mechanical oscillator and the atom are rapidly excited on the pulse arrival. In contrast, the excitation of photons is clearly delayed by a period of time. After a short-time interval according to the Gaussian envelope in Eq.~(\ref{gauss}), the driving intensity nearly vanishes, resulting in the subsequent dynamics being mainly determined by the full Hamiltonian~(\ref{fullH}) or the effective Hamiltonian~(\ref{H_eff}). The populations of the phonons and atom and the mean number of photons therefore manifest a complementary pattern in a sinusoidal-like evolution. In other words, the phonon number of the oscillator is in phase with the atomic excitation number and they are out of phase with the photon number. Confirmed by the Fourier analysis in Fig.~\ref{dynamic}(b), it is found that the excitation or energy transfer between the photon and the atom plus the phonon can be roughly described by the effective Hamiltonian
\begin{equation}\label{omega1}
H_{\rm eff}=g_{\rm eff}\left(|e01\rangle\langle g10|+|g10\rangle\langle e01|\right),
\end{equation}
with a transition rate
\begin{equation}
\omega_1=|g_{\rm eff}|=g\lambda\left(\frac{1}{2\omega_c-\omega_m}+\frac{1}{\omega_m}\right).
\end{equation}
It is evident that a peak around $\omega_1$ is distinguished in the frequency domain.

As the driving intensity $A$ increases, the mechanical oscillator can be populated to higher Fock states. The population dynamics in Figs.~\ref{dynamic}(c) and~\ref{dynamic}(e) then deviates gradually from the sinusoidal-like oscillation. This means that the dynamics becomes involved with oscillations of other frequencies. In regard to the photon generation, an extra three-wave-mixing process with higher Fock states, e.g., $|e02\rangle$ and $|e11\rangle$, can appear in addition to that in Eq.~(\ref{omega1}).

When $A=2\pi/3$, Fig.~\ref{dynamic}(d) suggests that the transitions $|g20\rangle\leftrightarrow|e11\rangle$ and $|g11\rangle\leftrightarrow|e02\rangle$ of the same transition frequency $\omega_2$ become significant in the dynamics. These two photon-generation channels or Rabi oscillations can be described by the effective Hamiltonian as
\begin{subequations}
\begin{align}
H_{\rm eff}&\approx \omega_2\left(|e11\rangle\langle g20|+|g20\rangle\langle e11|\right),\label{e11}\\
H_{\rm eff}&\approx \omega_2\left(|e02\rangle\langle g11|+|g11\rangle\langle e02|\right),\label{e02}
\end{align}
\end{subequations}
where $\omega_2=\sqrt{2}\omega_1$ can be obtained by Eq.~(\ref{coupling strength}) with $n=1$ and $m=0$ or with $n=0$ and $m=1$. In comparison to Fig.~\ref{dynamic}(b), an extra peak is present around $\omega_2$ in Fig.~\ref{dynamic}(d). A non-negligible distance exists between the numerical simulation and the analytical result for $\omega_2$, which is partially due to the fact that the energy shift in Eq.~(\ref{frequency shift}) is state dependent. The parametric setting for $|g10\rangle\leftrightarrow|e01\rangle$ in the effective Hamiltonian~(\ref{omega1}) is thus not suitable for Eqs.~(\ref{e11}) and (\ref{e02}). It renders a practical oscillation frequency larger than $\omega_2$. This discussion also applies to the next peak frequency $\omega_3$ in Fig.~\ref{dynamic}(f).

When the driving intensity is as strong as $A=\pi$, the Rabi oscillations $|g21\rangle\leftrightarrow|e12\rangle$ ($|g13\rangle\leftrightarrow|e04\rangle$ and $|g40\rangle\leftrightarrow|e31\rangle$) become distinguished in the population dynamics, although the transitions between lower Fock states still dominate in Fig.~\ref{dynamic}(f). In comparison to Figs.~\ref{dynamic}(b) and \ref{dynamic}(d), the extra oscillations in Fig.~\ref{dynamic}(f) are characterized by the transition frequency $\omega_3=2\omega_1$, which can be obtained by Eq.~(\ref{coupling strength}) with $n=1 (0 and 3)$ and $m=1 (3 and 0)$.

\subsection{Weak-coupling regime}\label{Weak-coupling regime}

\begin{figure}[htbp]
\centering
\includegraphics[width=0.95\linewidth]{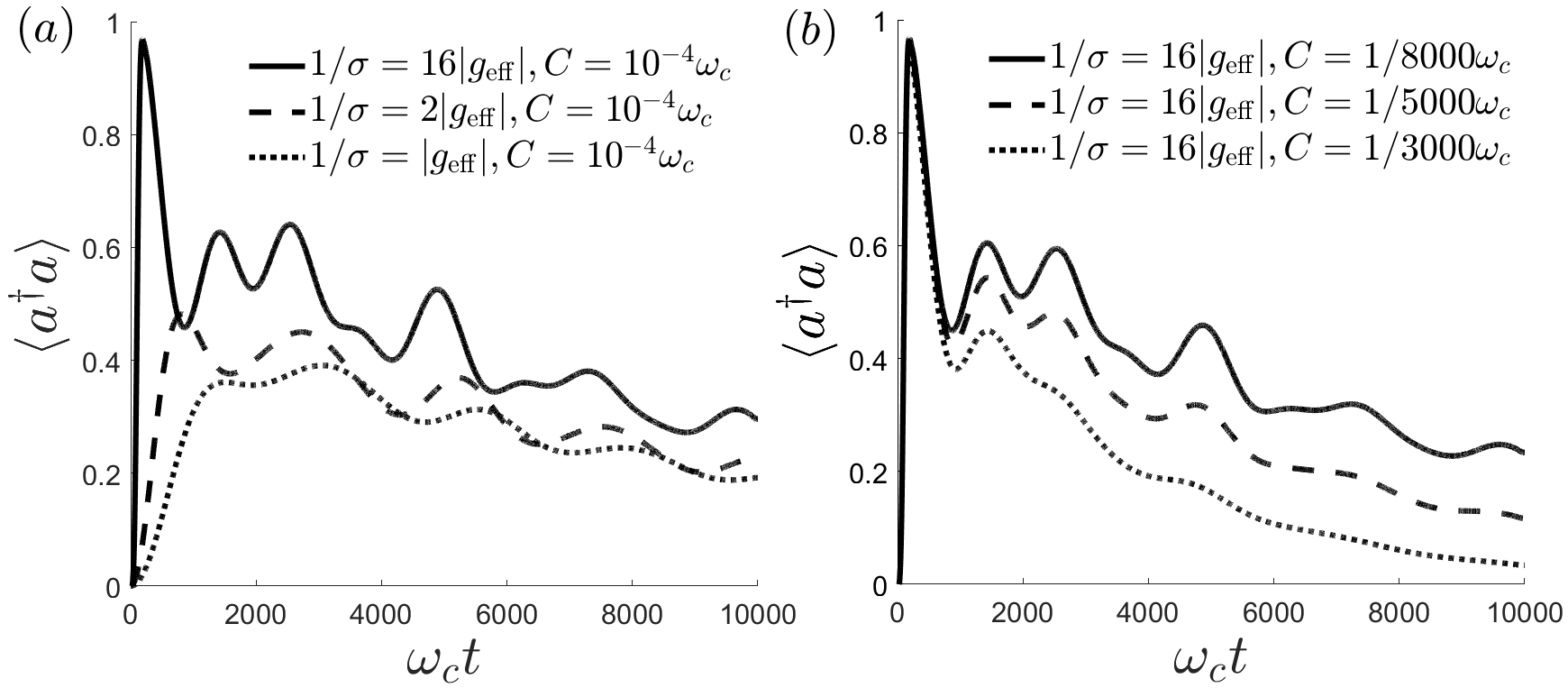}
\caption{Time evolution of the average population of the cavity mode under the pulse in Eq.~(15) with an intensity $A=\pi$. In (a), the loss rate is fixed as $C=10^{-4}\omega_c$ and the black solid, dashed, and dotted lines show the evolution with the standard deviations $1/\sigma=16|g_{\rm eff}|$, $2|g_{\rm eff}|$, and $|g_{\rm eff}|$, respectively. In (b), the standard deviation is fixed as $1/\sigma=16|g_{\rm eff}|$ and the black solid, dashed, and dotted lines show the evolution with the loss rates $C=1/8000\omega_c$, $1/5000\omega_c$, and $1/3000\omega_c$, respectively. }\label{intermediate_photon}
\end{figure}

In Sec.~\ref{Strong-coupling regime} we have demonstrated that the three-wave-mixing process by the effective Hamiltonian~(\ref{H_eff}) constitutes the main component during the system dynamics as well as the mechanism of converting the energy of the mechanical oscillator and atomic excitation into photons. The entire pattern of system dynamics depends on the time expansion $\sigma$ of the pulse envelope in Eq.~(\ref{ultrafast}), as well as the loss rates in the master equation~(\ref{master equation}). In Fig.~\ref{intermediate_photon} one can see a progressive change in dynamical pattern about the average photon number under various $\sigma$ and $C$. In Fig.~\ref{intermediate_photon}(a), a larger $\sigma$ yields a longer duration of the driving, resulting in more time for the photons to attain the peak value. In Fig.~\ref{intermediate_photon}(b) a larger $C$ induces a weaker oscillation during the photon generation. As $\sigma$ approaches infinity and the loss rate $C$ becomes greater than $|g_{\rm eff}|$, the strong-coupling regime moves to the weak-coupling regime and the combined effect from $\sigma$ and $C$ is expected to yield a stable output of photons.

In the weak-coupling regime, the driving functions in Eq.~(\ref{driving}) are modified to be
\begin{equation}\label{longdrive}
\begin{aligned}
  F_1(t)&=A\gamma\cos(\omega_m t), \\
  F_2(t)&=A\gamma\cos(\omega_a t),
\end{aligned}
\end{equation}
where the driving intensity is set as $A=12$, to continuously produce photons. In addition, we set the loss rate in the master equation~(\ref{master equation}) as $C=\kappa=\eta=10\gamma=2.5\times10^{-3}\omega_c$ to ensure that the effective coupling strength is significantly lower than the system loss, i.e., $|g_{\rm eff}|/C<1$.

\begin{figure}[htbp]
\centering
\includegraphics[width=0.95\linewidth]{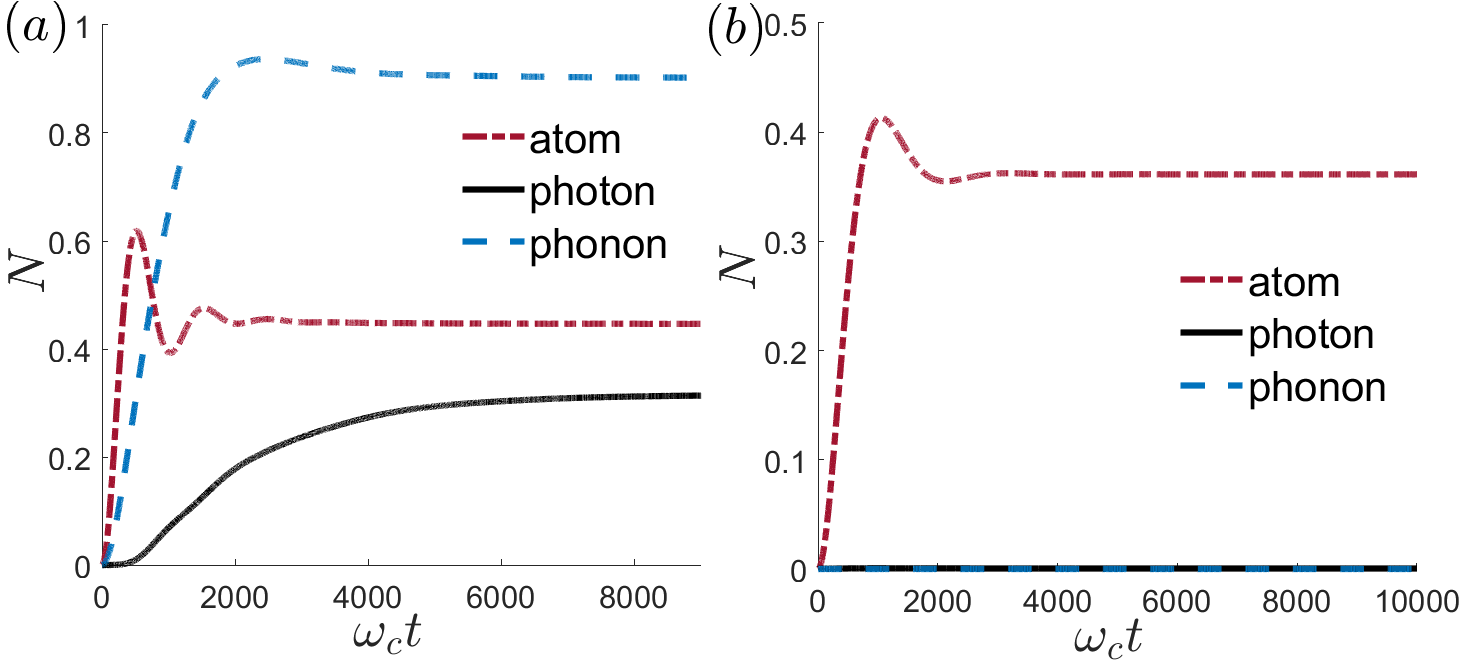}
\caption{Time evolution of the average population of the atom (red dot-dashed line), the mechanical oscillator (blue dashed line), and the cavity mode (black solid line) under the driving in Eq.~(\ref{longdrive}) when (a) both the mechanical oscillator and the atom are under driving and (b) only the atomic driving is on.}\label{MDCE}
\end{figure}

Figure~\ref{MDCE}(a) shows that the average populations of both the mechanical oscillator and atom rapidly rise under the continuous driving. After a time delay, the photon population starts increasing with a comparatively low rate. This can be inferred by the fact that the photon excitation comes from the phonon and atom via the three-wave mixing process after the phonon and atom are sufficiently populated through external driving. By definition, the MDCE can be observed after $\omega_ct\approx 5000$ when the three components approach steady states, as a balance between the two drivings and system dissipation. The photon population in the cavity is stabilized to be $0.31$ in Fig.~\ref{MDCE}(a), much larger than the result in Ref.~\cite{qin2019Emission}. A detuned two-photon driving was employed to squeeze the cavity mode in the optomechanical system~\cite{qin2019Emission}. Then the coupling between the mechanical oscillator and the cavity mode leads to an output flux of about $2.0\times10^4$ photons per second for a cavity with a typical linewidth of $\gamma/2\pi=2$ MHz. With the same leaking rate, our method can emit about $\gamma\langle a^\dagger a\rangle=3.8\times 10^6$ photons per second. This radiation can be measured using single-photon detectors~\cite{korzh2020Demonstration}. In current experiments, one can achieve the detection of individual microwave photons of a few gigahertz through techniques of quantum non-demolition measurement~\cite{rettaroli2025Novel}. To confirm that the photons in the cavity are practically generated by the vibration of the mechanical oscillator, we switched off the mechanical driving in Fig.~\ref{MDCE}(b). The population dynamics of the three components shows that the average photon number in the steady state nearly vanishes.

\begin{figure}[htbp]
\centering
\includegraphics[width=0.95\linewidth]{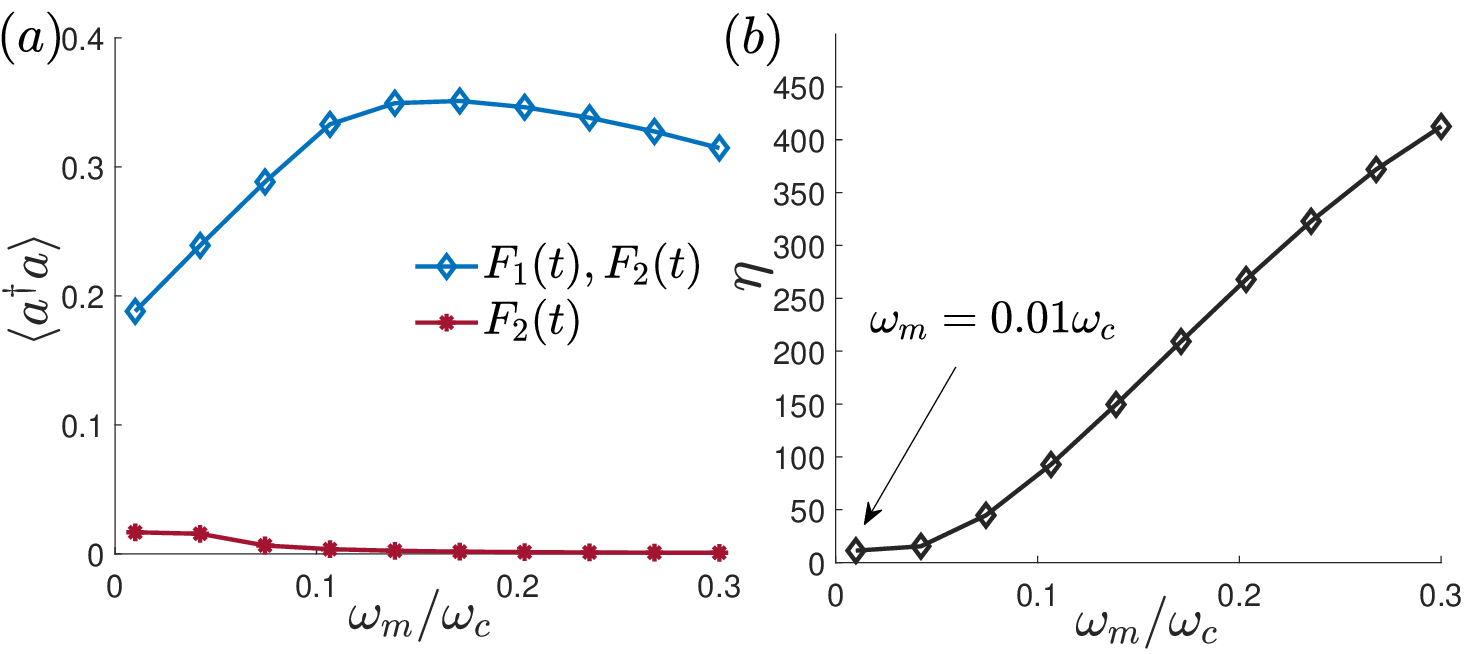}
\caption{(a) Stable photon number in the cavity as a function of the mechanical frequency under both atomic and mechanical driving (blue solid line with diamonds) or only under atomic driving (red solid line with asterisks). (b) Signal-to-noise ratio between the photons triggered by the three-wave-mixing process and those generated by the photon-atom Rabi oscillation as a function of the mechanical frequency.}\label{ratio}
\end{figure}

Note that the parametric setting in Fig.~\ref{MDCE} is the same as in Fig.~\ref{anticorossing}(a), where the mechanical frequency is about one order smaller than the output photon, i.e., $\omega_m=0.3\omega_c$. In practice, our method for realizing the MDCE can be extended to a wide range of low-frequency mechanical oscillators. In Fig.~\ref{ratio}(a) we present the photon population of the steady state from $\omega_m=0.01\omega_c$ to $0.3\omega_c$ obtained from either simultaneous driving of both the mechanical oscillator and qubit or driving only the qubit. These two cases demonstrate a dramatic distinction. Through driving both components, the excited phonon and atomic excitation can be constantly transferred to photons, reaching a detectable level of output. The stable population of photons in the cavity is not very sensitive to $\omega_m$. It is about $0.19$ when $\omega_m$ is as small as $0.01\omega_c$. Then it increases gradually with the frequency of the mechanical oscillator, until it attains a peak value of about $0.35$ around $\omega_m\approx0.15\omega_c$. This means that in such a wide range of $\omega_m$, the three-wave-mixing process remains the basic mechanism underlying the system dynamics, as long as the atomic frequency can be adjusted to meet the condition $\omega_c\approx\omega_a+\omega_m$. However, one can also find a slight photon generation by driving only the qubit, when $\omega_m$ decreases to be less than about $0.07\omega_c$ (see the red solid line with asterisks). This might be interpreted as a direct conversion between photon and atomic excitation, constituting a second resource of photons in addition to the three-wave mixing. A subsequent question is whether or not the result is still a MDCE when we employ a low-frequency mechanical oscillator in this hybrid system.

In Fig.~\ref{ratio}(b) a signal-to-noise ratio $\eta\equiv N_1/N_2$ is evaluated to identify the actual source of the photons, where $N_1$ and $N_2$ denote the steady-state numbers of signal photons and noisy photons generated by the three-wave-mixing process involving the mechanical oscillator and through the Rabi oscillation between the atom and the cavity mode, respectively. The ratio is found to monotonically increase with the mechanical frequency. Even when $\omega_m=0.01\omega_c$, $\eta$ is still over $11$. Therefore, it is confirmed that our method allows us to realize the MDCE with a low-frequency mechanical oscillator, which is at least two orders smaller than the previous method~\cite{macri2018Nonperturbative} based on a multiple-phonon process $k\omega_m=2\omega_c$. Our method is thus accessible in current experiments; see, e.g., Ref.~\cite{pirkkalainen2015Cavity}, where the mechanical oscillator frequency is about $\omega_m/2\pi\sim 65$ MHz, the cavity frequency $\omega_c/2\pi\sim 5$ GHz, and the radiation-pressure coupling between photons and phonons is up to $100$ MHz. Also, the resonant driving on both the mechanical oscillator and qubit has been realized, for example, in Ref.~\cite{oconnell2010Quantum}.

\section{Conclusion}\label{conclusion}

In this work, we have proposed a method for realizing instead of simulating the dynamical Casimir effect within a hybrid qubit-photon-phonon system. Our method yields appreciable photons in the cavity optomechanical system with a low-frequency mechanical oscillator, which greatly reduces the experimental cost in current platforms. Through an effective Hamiltonian based on the quantum-mechanical description of both the cavity field and the movable mirror, we identified that the three-wave-mixing process inherent in the system dynamics constitutes the fundamental mechanism for the mechanical DCE in our optomechanical system. The use of a frequency-tunable qubit overcomes the current technical limits for mechanical resonators. The relevant transitions for creating photons through deexciting the atom and mechanical oscillator provide a pathway for detecting the MDCE in the steady state under continuous driving and weak dissipation. We also used a signal-to-noise ratio to confirm that the output photon is really a consequence of the MDCE rather than from the Rabi oscillation between the qubit and photon. Our work appears to be a dramatic generation of the mechanical DCE in the low-frequency regime for the mechanical oscillator.

\bibliographystyle{apsrevlong}

\bibliography{ref}

\begin{thebibliography}{52}%
\makeatletter
\providecommand \@ifxundefined [1]{%
 \@ifx{#1\undefined}
}%
\providecommand \@ifnum [1]{%
 \ifnum #1\expandafter \@firstoftwo
 \else \expandafter \@secondoftwo
 \fi
}%
\providecommand \@ifx [1]{%
 \ifx #1\expandafter \@firstoftwo
 \else \expandafter \@secondoftwo
 \fi
}%
\providecommand \natexlab [1]{#1}%
\providecommand \enquote  [1]{``#1''}%
\providecommand \bibnamefont  [1]{#1}%
\providecommand \bibfnamefont [1]{#1}%
\providecommand \citenamefont [1]{#1}%
\providecommand \href@noop [0]{\@secondoftwo}%
\providecommand \href [0]{\begingroup \@sanitize@url \@href}%
\providecommand \@href[1]{\@@startlink{#1}\@@href}%
\providecommand \@@href[1]{\endgroup#1\@@endlink}%
\providecommand \@sanitize@url [0]{\catcode `\\12\catcode `\$12\catcode
  `\&12\catcode `\#12\catcode `\^12\catcode `\_12\catcode `\%12\relax}%
\providecommand \@@startlink[1]{}%
\providecommand \@@endlink[0]{}%
\providecommand \url  [0]{\begingroup\@sanitize@url \@url }%
\providecommand \@url [1]{\endgroup\@href {#1}{\urlprefix }}%
\providecommand \urlprefix  [0]{URL }%
\providecommand \Eprint [0]{\href }%
\providecommand \doibase [0]{http://dx.doi.org/}%
\providecommand \selectlanguage [0]{\@gobble}%
\providecommand \bibinfo  [0]{\@secondoftwo}%
\providecommand \bibfield  [0]{\@secondoftwo}%
\providecommand \translation [1]{[#1]}%
\providecommand \BibitemOpen [0]{}%
\providecommand \bibitemStop [0]{}%
\providecommand \bibitemNoStop [0]{.\EOS\space}%
\providecommand \EOS [0]{\spacefactor3000\relax}%
\providecommand \BibitemShut  [1]{\csname bibitem#1\endcsname}%
\let\auto@bib@innerbib\@empty
\bibitem [{\citenamefont {Hawking}(1974)}]{hawking1974Black}%
  \BibitemOpen
  \bibfield  {author} {\bibinfo {author} {\bibfnamefont {S.~W.}\ \bibnamefont
  {Hawking}},\ }\bibfield  {title} {\emph {\bibinfo {title} {Black hole
  explosions?}\ }}\href {\doibase 10.1038/248030a0} {\bibfield  {journal}
  {\bibinfo  {journal} {Nature (London)}\ }\textbf {\bibinfo {volume} {248}},\
  \bibinfo {pages} {30} (\bibinfo {year} {1974})}\BibitemShut {NoStop}%
\bibitem [{\citenamefont {Hawking}(1975)}]{hawking1975Particle}%
  \BibitemOpen
  \bibfield  {author} {\bibinfo {author} {\bibfnamefont {S.~W.}\ \bibnamefont
  {Hawking}},\ }\bibfield  {title} {\emph {\bibinfo {title} {Particle creation
  by black holes},\ }}\href {\doibase 10.1007/BF02345020} {\bibfield  {journal}
  {\bibinfo  {journal} {Commun.Math. Phys.}\ }\textbf {\bibinfo {volume}
  {43}},\ \bibinfo {pages} {199} (\bibinfo {year} {1975})}\BibitemShut
  {NoStop}%
\bibitem [{\citenamefont {Unruh}(1976)}]{unruh1976Notes}%
  \BibitemOpen
  \bibfield  {author} {\bibinfo {author} {\bibfnamefont {W.~G.}\ \bibnamefont
  {Unruh}},\ }\bibfield  {title} {\emph {\bibinfo {title} {Notes on black-hole
  evaporation},\ }}\href {\doibase 10.1103/PhysRevD.14.870} {\bibfield
  {journal} {\bibinfo  {journal} {Phys. Rev. D}\ }\textbf {\bibinfo {volume}
  {14}},\ \bibinfo {pages} {870} (\bibinfo {year} {1976})}\BibitemShut
  {NoStop}%
\bibitem [{\citenamefont {Wilson}\ \emph {et~al.}(2011)\citenamefont {Wilson},
  \citenamefont {Johansson}, \citenamefont {Pourkabirian}, \citenamefont
  {Simoen}, \citenamefont {Johansson}, \citenamefont {Duty}, \citenamefont
  {Nori},\ and\ \citenamefont {Delsing}}]{wilson2011Observation}%
  \BibitemOpen
  \bibfield  {author} {\bibinfo {author} {\bibfnamefont {C.~M.}\ \bibnamefont
  {Wilson}}, \bibinfo {author} {\bibfnamefont {G.}~\bibnamefont {Johansson}},
  \bibinfo {author} {\bibfnamefont {A.}~\bibnamefont {Pourkabirian}}, \bibinfo
  {author} {\bibfnamefont {M.}~\bibnamefont {Simoen}}, \bibinfo {author}
  {\bibfnamefont {J.~R.}\ \bibnamefont {Johansson}}, \bibinfo {author}
  {\bibfnamefont {T.}~\bibnamefont {Duty}}, \bibinfo {author} {\bibfnamefont
  {F.}~\bibnamefont {Nori}}, \ and\ \bibinfo {author} {\bibfnamefont
  {P.}~\bibnamefont {Delsing}},\ }\bibfield  {title} {\emph {\bibinfo {title}
  {Observation of the dynamical casimir effect in a superconducting circuit},\
  }}\href {\doibase 10.1038/nature10561} {\bibfield  {journal} {\bibinfo
  {journal} {Nature (London)}\ }\textbf {\bibinfo {volume} {479}},\ \bibinfo
  {pages} {376} (\bibinfo {year} {2011})}\BibitemShut {NoStop}%
\bibitem [{\citenamefont {Nation}\ \emph {et~al.}(2012)\citenamefont {Nation},
  \citenamefont {Johansson}, \citenamefont {Blencowe},\ and\ \citenamefont
  {Nori}}]{nation2012Colloquium}%
  \BibitemOpen
  \bibfield  {author} {\bibinfo {author} {\bibfnamefont {P.~D.}\ \bibnamefont
  {Nation}}, \bibinfo {author} {\bibfnamefont {J.~R.}\ \bibnamefont
  {Johansson}}, \bibinfo {author} {\bibfnamefont {M.~P.}\ \bibnamefont
  {Blencowe}}, \ and\ \bibinfo {author} {\bibfnamefont {F.}~\bibnamefont
  {Nori}},\ }\bibfield  {title} {\emph {\bibinfo {title} {{\emph{Colloquium}} :
  Stimulating uncertainty: Amplifying the quantum vacuum with superconducting
  circuits},\ }}\href {\doibase 10.1103/RevModPhys.84.1} {\bibfield  {journal}
  {\bibinfo  {journal} {Rev. Mod. Phys.}\ }\textbf {\bibinfo {volume} {84}},\
  \bibinfo {pages} {1} (\bibinfo {year} {2012})}\BibitemShut {NoStop}%
\bibitem [{\citenamefont {Benenti}\ \emph {et~al.}(2014)\citenamefont
  {Benenti}, \citenamefont {D'Arrigo}, \citenamefont {Siccardi},\ and\
  \citenamefont {Strini}}]{benenti2014Dynamical}%
  \BibitemOpen
  \bibfield  {author} {\bibinfo {author} {\bibfnamefont {G.}~\bibnamefont
  {Benenti}}, \bibinfo {author} {\bibfnamefont {A.}~\bibnamefont {D'Arrigo}},
  \bibinfo {author} {\bibfnamefont {S.}~\bibnamefont {Siccardi}}, \ and\
  \bibinfo {author} {\bibfnamefont {G.}~\bibnamefont {Strini}},\ }\bibfield
  {title} {\emph {\bibinfo {title} {Dynamical casimir effect in
  quantum-information processing},\ }}\href {\doibase
  10.1103/PhysRevA.90.052313} {\bibfield  {journal} {\bibinfo  {journal} {Phys.
  Rev. A}\ }\textbf {\bibinfo {volume} {90}},\ \bibinfo {pages} {052313}
  (\bibinfo {year} {2014})}\BibitemShut {NoStop}%
\bibitem [{\citenamefont {Macr{\`i}}\ \emph
  {et~al.}(2018{\natexlab{a}})\citenamefont {Macr{\`i}}, \citenamefont
  {Ridolfo}, \citenamefont {Di~Stefano}, \citenamefont {Kockum}, \citenamefont
  {Nori},\ and\ \citenamefont {Savasta}}]{macri2018Nonperturbative}%
  \BibitemOpen
  \bibfield  {author} {\bibinfo {author} {\bibfnamefont {V.}~\bibnamefont
  {Macr{\`i}}}, \bibinfo {author} {\bibfnamefont {A.}~\bibnamefont {Ridolfo}},
  \bibinfo {author} {\bibfnamefont {O.}~\bibnamefont {Di~Stefano}}, \bibinfo
  {author} {\bibfnamefont {A.~F.}\ \bibnamefont {Kockum}}, \bibinfo {author}
  {\bibfnamefont {F.}~\bibnamefont {Nori}}, \ and\ \bibinfo {author}
  {\bibfnamefont {S.}~\bibnamefont {Savasta}},\ }\bibfield  {title} {\emph
  {\bibinfo {title} {Nonperturbative dynamical casimir effect in optomechanical
  systems: Vacuum casimir-rabi splittings},\ }}\href {\doibase
  10.1103/PhysRevX.8.011031} {\bibfield  {journal} {\bibinfo  {journal} {Phys.
  Rev. X}\ }\textbf {\bibinfo {volume} {8}},\ \bibinfo {pages} {011031}
  (\bibinfo {year} {2018}{\natexlab{a}})}\BibitemShut {NoStop}%
\bibitem [{\citenamefont {Di~Stefano}\ \emph {et~al.}(2019)\citenamefont
  {Di~Stefano}, \citenamefont {Settineri}, \citenamefont {Macr{\`i}},
  \citenamefont {Ridolfo}, \citenamefont {Stassi}, \citenamefont {Kockum},
  \citenamefont {Savasta},\ and\ \citenamefont
  {Nori}}]{distefano2019Interaction}%
  \BibitemOpen
  \bibfield  {author} {\bibinfo {author} {\bibfnamefont {O.}~\bibnamefont
  {Di~Stefano}}, \bibinfo {author} {\bibfnamefont {A.}~\bibnamefont
  {Settineri}}, \bibinfo {author} {\bibfnamefont {V.}~\bibnamefont
  {Macr{\`i}}}, \bibinfo {author} {\bibfnamefont {A.}~\bibnamefont {Ridolfo}},
  \bibinfo {author} {\bibfnamefont {R.}~\bibnamefont {Stassi}}, \bibinfo
  {author} {\bibfnamefont {A.~F.}\ \bibnamefont {Kockum}}, \bibinfo {author}
  {\bibfnamefont {S.}~\bibnamefont {Savasta}}, \ and\ \bibinfo {author}
  {\bibfnamefont {F.}~\bibnamefont {Nori}},\ }\bibfield  {title} {\emph
  {\bibinfo {title} {Interaction of mechanical oscillators mediated by the
  exchange of virtual photon pairs},\ }}\href {\doibase
  10.1103/PhysRevLett.122.030402} {\bibfield  {journal} {\bibinfo  {journal}
  {Phys. Rev. Lett.}\ }\textbf {\bibinfo {volume} {122}},\ \bibinfo {pages}
  {030402} (\bibinfo {year} {2019})}\BibitemShut {NoStop}%
\bibitem [{\citenamefont {Casimir}(1948)}]{casimir1948Attraction}%
  \BibitemOpen
  \bibfield  {author} {\bibinfo {author} {\bibfnamefont {H.~B.~G.}\
  \bibnamefont {Casimir}},\ }\bibfield  {title} {\emph {\bibinfo {title} {On
  the attraction between two perfectly conducting plates},\ }}\href@noop {}
  {\bibfield  {journal} {\bibinfo  {journal} {Proc. Kon. Ned. Akad. Wet.}\
  }\textbf {\bibinfo {volume} {51}},\ \bibinfo {pages} {793} (\bibinfo {year}
  {1948})}\BibitemShut {NoStop}%
\bibitem [{\citenamefont {Sparnaay}(1958)}]{sparnaay1958Measurements}%
  \BibitemOpen
  \bibfield  {author} {\bibinfo {author} {\bibfnamefont {M.~J.}\ \bibnamefont
  {Sparnaay}},\ }\bibfield  {title} {\emph {\bibinfo {title} {Measurements of
  attractive forces between flat plates},\ }}\href {\doibase
  10.1016/S0031-8914(58)80090-7} {\bibfield  {journal} {\bibinfo  {journal}
  {Physica}\ }\textbf {\bibinfo {volume} {24}},\ \bibinfo {pages} {751}
  (\bibinfo {year} {1958})}\BibitemShut {NoStop}%
\bibitem [{\citenamefont {Lamoreaux}(1997)}]{lamoreaux1997Demonstration}%
  \BibitemOpen
  \bibfield  {author} {\bibinfo {author} {\bibfnamefont {S.~K.}\ \bibnamefont
  {Lamoreaux}},\ }\bibfield  {title} {\emph {\bibinfo {title} {Demonstration of
  the casimir force in the 0.6 to
  \$6{\textbackslash}ensuremath\{{\textbackslash}mu\}m\$ range},\ }}\href
  {\doibase 10.1103/PhysRevLett.78.5} {\bibfield  {journal} {\bibinfo
  {journal} {Phys. Rev. Lett.}\ }\textbf {\bibinfo {volume} {78}},\ \bibinfo
  {pages} {5} (\bibinfo {year} {1997})}\BibitemShut {NoStop}%
\bibitem [{\citenamefont {L{\"a}hteenm{\"a}ki}\ \emph
  {et~al.}(2013)\citenamefont {L{\"a}hteenm{\"a}ki}, \citenamefont {Paraoanu},
  \citenamefont {Hassel},\ and\ \citenamefont
  {Hakonen}}]{lahteenmaki2013Dynamical}%
  \BibitemOpen
  \bibfield  {author} {\bibinfo {author} {\bibfnamefont {P.}~\bibnamefont
  {L{\"a}hteenm{\"a}ki}}, \bibinfo {author} {\bibfnamefont {G.~S.}\
  \bibnamefont {Paraoanu}}, \bibinfo {author} {\bibfnamefont {J.}~\bibnamefont
  {Hassel}}, \ and\ \bibinfo {author} {\bibfnamefont {P.~J.}\ \bibnamefont
  {Hakonen}},\ }\bibfield  {title} {\emph {\bibinfo {title} {Dynamical casimir
  effect in a josephson metamaterial},\ }}\href {\doibase
  10.1073/pnas.1212705110} {\bibfield  {journal} {\bibinfo  {journal} {Proc.
  Natl. Acad. Sci. U.S.A.}\ }\textbf {\bibinfo {volume} {110}},\ \bibinfo
  {pages} {4234} (\bibinfo {year} {2013})}\BibitemShut {NoStop}%
\bibitem [{\citenamefont {Moore}(2003)}]{moore2003Quantum}%
  \BibitemOpen
  \bibfield  {author} {\bibinfo {author} {\bibfnamefont {G.~T.}\ \bibnamefont
  {Moore}},\ }\bibfield  {title} {\emph {\bibinfo {title} {Quantum theory of
  the electromagnetic field in a variable-length one-dimensional cavity},\
  }}\href {\doibase 10.1063/1.1665432} {\bibfield  {journal} {\bibinfo
  {journal} {J. Math. Phys.}\ }\textbf {\bibinfo {volume} {11}},\ \bibinfo
  {pages} {2679} (\bibinfo {year} {2003})}\BibitemShut {NoStop}%
\bibitem [{\citenamefont {Lambrecht}\ \emph {et~al.}(1996)\citenamefont
  {Lambrecht}, \citenamefont {Jaekel},\ and\ \citenamefont
  {Reynaud}}]{lambrecht1996Motion}%
  \BibitemOpen
  \bibfield  {author} {\bibinfo {author} {\bibfnamefont {A.}~\bibnamefont
  {Lambrecht}}, \bibinfo {author} {\bibfnamefont {M.-T.}\ \bibnamefont
  {Jaekel}}, \ and\ \bibinfo {author} {\bibfnamefont {S.}~\bibnamefont
  {Reynaud}},\ }\bibfield  {title} {\emph {\bibinfo {title} {Motion induced
  radiation from a vibrating cavity},\ }}\href {\doibase
  10.1103/PhysRevLett.77.615} {\bibfield  {journal} {\bibinfo  {journal} {Phys.
  Rev. Lett.}\ }\textbf {\bibinfo {volume} {77}},\ \bibinfo {pages} {615}
  (\bibinfo {year} {1996})}\BibitemShut {NoStop}%
\bibitem [{\citenamefont {Dodonov}(2020)}]{dodonov2020Fifty}%
  \BibitemOpen
  \bibfield  {author} {\bibinfo {author} {\bibfnamefont {V.}~\bibnamefont
  {Dodonov}},\ }\bibfield  {title} {\emph {\bibinfo {title} {Fifty years of the
  dynamical casimir effect},\ }}\href {\doibase 10.3390/physics2010007}
  {\bibfield  {journal} {\bibinfo  {journal} {Physics}\ }\textbf {\bibinfo
  {volume} {2}},\ \bibinfo {pages} {67} (\bibinfo {year} {2020})}\BibitemShut
  {NoStop}%
\bibitem [{\citenamefont {Qin}\ \emph {et~al.}(2019)\citenamefont {Qin},
  \citenamefont {Macr{\`i}}, \citenamefont {Miranowicz}, \citenamefont
  {Savasta},\ and\ \citenamefont {Nori}}]{qin2019Emission}%
  \BibitemOpen
  \bibfield  {author} {\bibinfo {author} {\bibfnamefont {W.}~\bibnamefont
  {Qin}}, \bibinfo {author} {\bibfnamefont {V.}~\bibnamefont {Macr{\`i}}},
  \bibinfo {author} {\bibfnamefont {A.}~\bibnamefont {Miranowicz}}, \bibinfo
  {author} {\bibfnamefont {S.}~\bibnamefont {Savasta}}, \ and\ \bibinfo
  {author} {\bibfnamefont {F.}~\bibnamefont {Nori}},\ }\bibfield  {title}
  {\emph {\bibinfo {title} {Emission of photon pairs by mechanical stimulation
  of the squeezed vacuum},\ }}\href {\doibase 10.1103/PhysRevA.100.062501}
  {\bibfield  {journal} {\bibinfo  {journal} {Phys. Rev. A}\ }\textbf {\bibinfo
  {volume} {100}},\ \bibinfo {pages} {062501} (\bibinfo {year}
  {2019})}\BibitemShut {NoStop}%
\bibitem [{\citenamefont {Brevik}\ \emph {et~al.}(2000)\citenamefont {Brevik},
  \citenamefont {Milton}, \citenamefont {Odintsov},\ and\ \citenamefont
  {Osetrin}}]{brevik2000Dynamical}%
  \BibitemOpen
  \bibfield  {author} {\bibinfo {author} {\bibfnamefont {I.}~\bibnamefont
  {Brevik}}, \bibinfo {author} {\bibfnamefont {K.~A.}\ \bibnamefont {Milton}},
  \bibinfo {author} {\bibfnamefont {S.~D.}\ \bibnamefont {Odintsov}}, \ and\
  \bibinfo {author} {\bibfnamefont {K.~E.}\ \bibnamefont {Osetrin}},\
  }\bibfield  {title} {\emph {\bibinfo {title} {Dynamical casimir effect and
  quantum cosmology},\ }}\href {\doibase 10.1103/PhysRevD.62.064005} {\bibfield
   {journal} {\bibinfo  {journal} {Phys. Rev. D}\ }\textbf {\bibinfo {volume}
  {62}},\ \bibinfo {pages} {064005} (\bibinfo {year} {2000})}\BibitemShut
  {NoStop}%
\bibitem [{\citenamefont {Ruser}\ and\ \citenamefont
  {Durrer}(2007)}]{ruser2007Dynamical}%
  \BibitemOpen
  \bibfield  {author} {\bibinfo {author} {\bibfnamefont {M.}~\bibnamefont
  {Ruser}}\ and\ \bibinfo {author} {\bibfnamefont {R.}~\bibnamefont {Durrer}},\
  }\bibfield  {title} {\emph {\bibinfo {title} {Dynamical casimir effect for
  gravitons in bouncing braneworlds},\ }}\href {\doibase
  10.1103/PhysRevD.76.104014} {\bibfield  {journal} {\bibinfo  {journal} {Phys.
  Rev. D}\ }\textbf {\bibinfo {volume} {76}},\ \bibinfo {pages} {104014}
  (\bibinfo {year} {2007})}\BibitemShut {NoStop}%
\bibitem [{\citenamefont {Lock}\ and\ \citenamefont
  {Fuentes}(2017)}]{lock2017Dynamical}%
  \BibitemOpen
  \bibfield  {author} {\bibinfo {author} {\bibfnamefont {M.~P.~E.}\
  \bibnamefont {Lock}}\ and\ \bibinfo {author} {\bibfnamefont {I.}~\bibnamefont
  {Fuentes}},\ }\bibfield  {title} {\emph {\bibinfo {title} {Dynamical casimir
  effect in curved spacetime},\ }}\href {\doibase 10.1088/1367-2630/aa7651}
  {\bibfield  {journal} {\bibinfo  {journal} {New J. Phys.}\ }\textbf {\bibinfo
  {volume} {19}},\ \bibinfo {pages} {073005} (\bibinfo {year}
  {2017})}\BibitemShut {NoStop}%
\bibitem [{\citenamefont {Ottewill}\ and\ \citenamefont
  {Takagi}(1988)}]{ottewill1988Radiation}%
  \BibitemOpen
  \bibfield  {author} {\bibinfo {author} {\bibfnamefont {A.}~\bibnamefont
  {Ottewill}}\ and\ \bibinfo {author} {\bibfnamefont {S.}~\bibnamefont
  {Takagi}},\ }\bibfield  {title} {\emph {\bibinfo {title} {Radiation by moving
  mirrors in curved space-time},\ }}\href {\doibase 10.1143/PTP.79.429}
  {\bibfield  {journal} {\bibinfo  {journal} {Prog. Theor. Phys.}\ }\textbf
  {\bibinfo {volume} {79}},\ \bibinfo {pages} {429} (\bibinfo {year}
  {1988})}\BibitemShut {NoStop}%
\bibitem [{\citenamefont {Ferreri}\ \emph {et~al.}(2024)\citenamefont
  {Ferreri}, \citenamefont {Bruschi}, \citenamefont {Wilhelm}, \citenamefont
  {Nori},\ and\ \citenamefont {Macr{\`i}}}]{ferreri2024Phononphoton}%
  \BibitemOpen
  \bibfield  {author} {\bibinfo {author} {\bibfnamefont {A.}~\bibnamefont
  {Ferreri}}, \bibinfo {author} {\bibfnamefont {D.~E.}\ \bibnamefont
  {Bruschi}}, \bibinfo {author} {\bibfnamefont {F.~K.}\ \bibnamefont
  {Wilhelm}}, \bibinfo {author} {\bibfnamefont {F.}~\bibnamefont {Nori}}, \
  and\ \bibinfo {author} {\bibfnamefont {V.}~\bibnamefont {Macr{\`i}}},\
  }\bibfield  {title} {\emph {\bibinfo {title} {Phonon-photon conversion as
  mechanism for cooling and coherence transfer},\ }}\href {\doibase
  10.1103/PhysRevResearch.6.023320} {\bibfield  {journal} {\bibinfo  {journal}
  {Phys. Rev. Res.}\ }\textbf {\bibinfo {volume} {6}},\ \bibinfo {pages}
  {023320} (\bibinfo {year} {2024})}\BibitemShut {NoStop}%
\bibitem [{\citenamefont {Ferreri}\ \emph {et~al.}(2023)\citenamefont
  {Ferreri}, \citenamefont {Macr{\`i}}, \citenamefont {Wilhelm}, \citenamefont
  {Nori},\ and\ \citenamefont {Bruschi}}]{ferreri2023Quantum}%
  \BibitemOpen
  \bibfield  {author} {\bibinfo {author} {\bibfnamefont {A.}~\bibnamefont
  {Ferreri}}, \bibinfo {author} {\bibfnamefont {V.}~\bibnamefont {Macr{\`i}}},
  \bibinfo {author} {\bibfnamefont {F.~K.}\ \bibnamefont {Wilhelm}}, \bibinfo
  {author} {\bibfnamefont {F.}~\bibnamefont {Nori}}, \ and\ \bibinfo {author}
  {\bibfnamefont {D.~E.}\ \bibnamefont {Bruschi}},\ }\bibfield  {title} {\emph
  {\bibinfo {title} {Quantum field heat engine powered by phonon-photon
  interactions},\ }}\href {\doibase 10.1103/PhysRevResearch.5.043274}
  {\bibfield  {journal} {\bibinfo  {journal} {Phys. Rev. Res.}\ }\textbf
  {\bibinfo {volume} {5}},\ \bibinfo {pages} {043274} (\bibinfo {year}
  {2023})}\BibitemShut {NoStop}%
\bibitem [{\citenamefont {Felicetti}\ \emph {et~al.}(2014)\citenamefont
  {Felicetti}, \citenamefont {Sanz}, \citenamefont {Lamata}, \citenamefont
  {Romero}, \citenamefont {Johansson}, \citenamefont {Delsing},\ and\
  \citenamefont {Solano}}]{felicetti2014Dynamical}%
  \BibitemOpen
  \bibfield  {author} {\bibinfo {author} {\bibfnamefont {S.}~\bibnamefont
  {Felicetti}}, \bibinfo {author} {\bibfnamefont {M.}~\bibnamefont {Sanz}},
  \bibinfo {author} {\bibfnamefont {L.}~\bibnamefont {Lamata}}, \bibinfo
  {author} {\bibfnamefont {G.}~\bibnamefont {Romero}}, \bibinfo {author}
  {\bibfnamefont {G.}~\bibnamefont {Johansson}}, \bibinfo {author}
  {\bibfnamefont {P.}~\bibnamefont {Delsing}}, \ and\ \bibinfo {author}
  {\bibfnamefont {E.}~\bibnamefont {Solano}},\ }\bibfield  {title} {\emph
  {\bibinfo {title} {Dynamical casimir effect entangles artificial atoms},\
  }}\href {\doibase 10.1103/PhysRevLett.113.093602} {\bibfield  {journal}
  {\bibinfo  {journal} {Phys. Rev. Lett.}\ }\textbf {\bibinfo {volume} {113}},\
  \bibinfo {pages} {093602} (\bibinfo {year} {2014})}\BibitemShut {NoStop}%
\bibitem [{\citenamefont {Busch}\ \emph {et~al.}(2014)\citenamefont {Busch},
  \citenamefont {Parentani},\ and\ \citenamefont
  {Robertson}}]{busch2014Quantum}%
  \BibitemOpen
  \bibfield  {author} {\bibinfo {author} {\bibfnamefont {X.}~\bibnamefont
  {Busch}}, \bibinfo {author} {\bibfnamefont {R.}~\bibnamefont {Parentani}}, \
  and\ \bibinfo {author} {\bibfnamefont {S.}~\bibnamefont {Robertson}},\
  }\bibfield  {title} {\emph {\bibinfo {title} {Quantum entanglement due to a
  modulated dynamical casimir effect},\ }}\href {\doibase
  10.1103/PhysRevA.89.063606} {\bibfield  {journal} {\bibinfo  {journal} {Phys.
  Rev. A}\ }\textbf {\bibinfo {volume} {89}},\ \bibinfo {pages} {063606}
  (\bibinfo {year} {2014})}\BibitemShut {NoStop}%
\bibitem [{\citenamefont {Aron}\ \emph {et~al.}(2014)\citenamefont {Aron},
  \citenamefont {Kulkarni},\ and\ \citenamefont
  {T{\"u}reci}}]{aron2014Steadystate}%
  \BibitemOpen
  \bibfield  {author} {\bibinfo {author} {\bibfnamefont {C.}~\bibnamefont
  {Aron}}, \bibinfo {author} {\bibfnamefont {M.}~\bibnamefont {Kulkarni}}, \
  and\ \bibinfo {author} {\bibfnamefont {H.~E.}\ \bibnamefont {T{\"u}reci}},\
  }\bibfield  {title} {\emph {\bibinfo {title} {Steady-state entanglement of
  spatially separated qubits via quantum bath engineering},\ }}\href {\doibase
  10.1103/PhysRevA.90.062305} {\bibfield  {journal} {\bibinfo  {journal} {Phys.
  Rev. A}\ }\textbf {\bibinfo {volume} {90}},\ \bibinfo {pages} {062305}
  (\bibinfo {year} {2014})}\BibitemShut {NoStop}%
\bibitem [{\citenamefont {Rossatto}\ \emph {et~al.}(2016)\citenamefont
  {Rossatto}, \citenamefont {Felicetti}, \citenamefont {Eneriz}, \citenamefont
  {Rico}, \citenamefont {Sanz},\ and\ \citenamefont
  {Solano}}]{rossatto2016Entangling}%
  \BibitemOpen
  \bibfield  {author} {\bibinfo {author} {\bibfnamefont {D.~Z.}\ \bibnamefont
  {Rossatto}}, \bibinfo {author} {\bibfnamefont {S.}~\bibnamefont {Felicetti}},
  \bibinfo {author} {\bibfnamefont {H.}~\bibnamefont {Eneriz}}, \bibinfo
  {author} {\bibfnamefont {E.}~\bibnamefont {Rico}}, \bibinfo {author}
  {\bibfnamefont {M.}~\bibnamefont {Sanz}}, \ and\ \bibinfo {author}
  {\bibfnamefont {E.}~\bibnamefont {Solano}},\ }\bibfield  {title} {\emph
  {\bibinfo {title} {Entangling polaritons via dynamical casimir effect in
  circuit quantum electrodynamics},\ }}\href {\doibase
  10.1103/PhysRevB.93.094514} {\bibfield  {journal} {\bibinfo  {journal} {Phys.
  Rev. B}\ }\textbf {\bibinfo {volume} {93}},\ \bibinfo {pages} {094514}
  (\bibinfo {year} {2016})}\BibitemShut {NoStop}%
\bibitem [{\citenamefont {Agust{\'i}}\ \emph {et~al.}(2019)\citenamefont
  {Agust{\'i}}, \citenamefont {Solano},\ and\ \citenamefont
  {Sab{\'i}n}}]{agusti2019Entanglement}%
  \BibitemOpen
  \bibfield  {author} {\bibinfo {author} {\bibfnamefont {A.}~\bibnamefont
  {Agust{\'i}}}, \bibinfo {author} {\bibfnamefont {E.}~\bibnamefont {Solano}},
  \ and\ \bibinfo {author} {\bibfnamefont {C.}~\bibnamefont {Sab{\'i}n}},\
  }\bibfield  {title} {\emph {\bibinfo {title} {Entanglement through qubit
  motion and the dynamical casimir effect},\ }}\href {\doibase
  10.1103/PhysRevA.99.052328} {\bibfield  {journal} {\bibinfo  {journal} {Phys.
  Rev. A}\ }\textbf {\bibinfo {volume} {99}},\ \bibinfo {pages} {052328}
  (\bibinfo {year} {2019})}\BibitemShut {NoStop}%
\bibitem [{\citenamefont {Friis}\ \emph {et~al.}(2012)\citenamefont {Friis},
  \citenamefont {Huber}, \citenamefont {Fuentes},\ and\ \citenamefont
  {Bruschi}}]{friis2012Quantum}%
  \BibitemOpen
  \bibfield  {author} {\bibinfo {author} {\bibfnamefont {N.}~\bibnamefont
  {Friis}}, \bibinfo {author} {\bibfnamefont {M.}~\bibnamefont {Huber}},
  \bibinfo {author} {\bibfnamefont {I.}~\bibnamefont {Fuentes}}, \ and\
  \bibinfo {author} {\bibfnamefont {D.~E.}\ \bibnamefont {Bruschi}},\
  }\bibfield  {title} {\emph {\bibinfo {title} {Quantum gates and multipartite
  entanglement resonances realized by nonuniform cavity motion},\ }}\href
  {\doibase 10.1103/PhysRevD.86.105003} {\bibfield  {journal} {\bibinfo
  {journal} {Phys. Rev. D}\ }\textbf {\bibinfo {volume} {86}},\ \bibinfo
  {pages} {105003} (\bibinfo {year} {2012})}\BibitemShut {NoStop}%
\bibitem [{\citenamefont {Bruschi}\ \emph {et~al.}(2013)\citenamefont
  {Bruschi}, \citenamefont {Dragan}, \citenamefont {Lee}, \citenamefont
  {Fuentes},\ and\ \citenamefont {Louko}}]{bruschi2013Relativistic}%
  \BibitemOpen
  \bibfield  {author} {\bibinfo {author} {\bibfnamefont {D.~E.}\ \bibnamefont
  {Bruschi}}, \bibinfo {author} {\bibfnamefont {A.}~\bibnamefont {Dragan}},
  \bibinfo {author} {\bibfnamefont {A.~R.}\ \bibnamefont {Lee}}, \bibinfo
  {author} {\bibfnamefont {I.}~\bibnamefont {Fuentes}}, \ and\ \bibinfo
  {author} {\bibfnamefont {J.}~\bibnamefont {Louko}},\ }\bibfield  {title}
  {\emph {\bibinfo {title} {Relativistic motion generates quantum gates and
  entanglement resonances},\ }}\href {\doibase 10.1103/PhysRevLett.111.090504}
  {\bibfield  {journal} {\bibinfo  {journal} {Phys. Rev. Lett.}\ }\textbf
  {\bibinfo {volume} {111}},\ \bibinfo {pages} {090504} (\bibinfo {year}
  {2013})}\BibitemShut {NoStop}%
\bibitem [{\citenamefont {Sab{\'i}n}\ and\ \citenamefont
  {Adesso}(2015)}]{sabin2015Generation}%
  \BibitemOpen
  \bibfield  {author} {\bibinfo {author} {\bibfnamefont {C.}~\bibnamefont
  {Sab{\'i}n}}\ and\ \bibinfo {author} {\bibfnamefont {G.}~\bibnamefont
  {Adesso}},\ }\bibfield  {title} {\emph {\bibinfo {title} {Generation of
  quantum steering and interferometric power in the dynamical casimir effect},\
  }}\href {\doibase 10.1103/PhysRevA.92.042107} {\bibfield  {journal} {\bibinfo
   {journal} {Phys. Rev. A}\ }\textbf {\bibinfo {volume} {92}},\ \bibinfo
  {pages} {042107} (\bibinfo {year} {2015})}\BibitemShut {NoStop}%
\bibitem [{\citenamefont {O'Connell}\ \emph {et~al.}(2010)\citenamefont
  {O'Connell}, \citenamefont {Hofheinz}, \citenamefont {Ansmann},\ and\
  \citenamefont {Bialczak}}]{oconnell2010Quantum}%
  \BibitemOpen
  \bibfield  {author} {\bibinfo {author} {\bibfnamefont {A.~D.}\ \bibnamefont
  {O'Connell}}, \bibinfo {author} {\bibfnamefont {M.}~\bibnamefont {Hofheinz}},
  \bibinfo {author} {\bibfnamefont {M.}~\bibnamefont {Ansmann}}, \ and\
  \bibinfo {author} {\bibfnamefont {R.~C.}\ \bibnamefont {Bialczak}},\
  }\bibfield  {title} {\emph {\bibinfo {title} {Quantum ground state and
  single-phonon control of a mechanical resonator},\ }}\href {\doibase
  10.1038/nature08967} {\bibfield  {journal} {\bibinfo  {journal} {Nature
  (London)}\ }\textbf {\bibinfo {volume} {464}},\ \bibinfo {pages} {697}
  (\bibinfo {year} {2010})}\BibitemShut {NoStop}%
\bibitem [{\citenamefont {Johansson}\ \emph {et~al.}(2009)\citenamefont
  {Johansson}, \citenamefont {Johansson}, \citenamefont {Wilson},\ and\
  \citenamefont {Nori}}]{johansson2009Dynamical}%
  \BibitemOpen
  \bibfield  {author} {\bibinfo {author} {\bibfnamefont {J.~R.}\ \bibnamefont
  {Johansson}}, \bibinfo {author} {\bibfnamefont {G.}~\bibnamefont
  {Johansson}}, \bibinfo {author} {\bibfnamefont {C.~M.}\ \bibnamefont
  {Wilson}}, \ and\ \bibinfo {author} {\bibfnamefont {F.}~\bibnamefont
  {Nori}},\ }\bibfield  {title} {\emph {\bibinfo {title} {Dynamical casimir
  effect in a superconducting coplanar waveguide},\ }}\href {\doibase
  10.1103/PhysRevLett.103.147003} {\bibfield  {journal} {\bibinfo  {journal}
  {Phys. Rev. Lett.}\ }\textbf {\bibinfo {volume} {103}},\ \bibinfo {pages}
  {147003} (\bibinfo {year} {2009})}\BibitemShut {NoStop}%
\bibitem [{\citenamefont {Johansson}\ \emph {et~al.}(2010)\citenamefont
  {Johansson}, \citenamefont {Johansson}, \citenamefont {Wilson},\ and\
  \citenamefont {Nori}}]{johansson2010Dynamical}%
  \BibitemOpen
  \bibfield  {author} {\bibinfo {author} {\bibfnamefont {J.~R.}\ \bibnamefont
  {Johansson}}, \bibinfo {author} {\bibfnamefont {G.}~\bibnamefont
  {Johansson}}, \bibinfo {author} {\bibfnamefont {C.~M.}\ \bibnamefont
  {Wilson}}, \ and\ \bibinfo {author} {\bibfnamefont {F.}~\bibnamefont
  {Nori}},\ }\bibfield  {title} {\emph {\bibinfo {title} {Dynamical casimir
  effect in superconducting microwave circuits},\ }}\href {\doibase
  10.1103/PhysRevA.82.052509} {\bibfield  {journal} {\bibinfo  {journal} {Phys.
  Rev. A}\ }\textbf {\bibinfo {volume} {82}},\ \bibinfo {pages} {052509}
  (\bibinfo {year} {2010})}\BibitemShut {NoStop}%
\bibitem [{\citenamefont {Crocce}\ \emph {et~al.}(2004)\citenamefont {Crocce},
  \citenamefont {Dalvit}, \citenamefont {Lombardo},\ and\ \citenamefont
  {Mazzitelli}}]{crocce2004Model}%
  \BibitemOpen
  \bibfield  {author} {\bibinfo {author} {\bibfnamefont {M.}~\bibnamefont
  {Crocce}}, \bibinfo {author} {\bibfnamefont {D.~A.~R.}\ \bibnamefont
  {Dalvit}}, \bibinfo {author} {\bibfnamefont {F.~C.}\ \bibnamefont
  {Lombardo}}, \ and\ \bibinfo {author} {\bibfnamefont {F.~D.}\ \bibnamefont
  {Mazzitelli}},\ }\bibfield  {title} {\emph {\bibinfo {title} {Model for
  resonant photon creation in a cavity with time-dependent conductivity},\
  }}\href {\doibase 10.1103/PhysRevA.70.033811} {\bibfield  {journal} {\bibinfo
   {journal} {Phys. Rev. A}\ }\textbf {\bibinfo {volume} {70}},\ \bibinfo
  {pages} {033811} (\bibinfo {year} {2004})}\BibitemShut {NoStop}%
\bibitem [{\citenamefont {Dodonov}(2010)}]{dodonov2010Current}%
  \BibitemOpen
  \bibfield  {author} {\bibinfo {author} {\bibfnamefont {V.~V.}\ \bibnamefont
  {Dodonov}},\ }\bibfield  {title} {\emph {\bibinfo {title} {Current status of
  the dynamical casimir effect},\ }}\href {\doibase
  10.1088/0031-8949/82/03/038105} {\bibfield  {journal} {\bibinfo  {journal}
  {Phys. Scr.}\ }\textbf {\bibinfo {volume} {82}},\ \bibinfo {pages} {038105}
  (\bibinfo {year} {2010})}\BibitemShut {NoStop}%
\bibitem [{\citenamefont {Kim}\ \emph {et~al.}(2006)\citenamefont {Kim},
  \citenamefont {Brownell},\ and\ \citenamefont
  {Onofrio}}]{kim2006Detectability}%
  \BibitemOpen
  \bibfield  {author} {\bibinfo {author} {\bibfnamefont {W.-J.}\ \bibnamefont
  {Kim}}, \bibinfo {author} {\bibfnamefont {J.~H.}\ \bibnamefont {Brownell}}, \
  and\ \bibinfo {author} {\bibfnamefont {R.}~\bibnamefont {Onofrio}},\
  }\bibfield  {title} {\emph {\bibinfo {title} {Detectability of dissipative
  motion in quantum vacuum via superradiance},\ }}\href {\doibase
  10.1103/PhysRevLett.96.200402} {\bibfield  {journal} {\bibinfo  {journal}
  {Phys. Rev. Lett.}\ }\textbf {\bibinfo {volume} {96}},\ \bibinfo {pages}
  {200402} (\bibinfo {year} {2006})}\BibitemShut {NoStop}%
\bibitem [{\citenamefont {Brownell}\ \emph {et~al.}(2008)\citenamefont
  {Brownell}, \citenamefont {Kim},\ and\ \citenamefont
  {Onofrio}}]{brownell2008Modelling}%
  \BibitemOpen
  \bibfield  {author} {\bibinfo {author} {\bibfnamefont {J.~H.}\ \bibnamefont
  {Brownell}}, \bibinfo {author} {\bibfnamefont {W.~J.}\ \bibnamefont {Kim}}, \
  and\ \bibinfo {author} {\bibfnamefont {R.}~\bibnamefont {Onofrio}},\
  }\bibfield  {title} {\emph {\bibinfo {title} {Modelling superradiant
  amplification of casimir photons in very low dissipation cavities},\ }}\href
  {\doibase 10.1088/1751-8113/41/16/164026} {\bibfield  {journal} {\bibinfo
  {journal} {J. Phys. A: Math. Theor.}\ }\textbf {\bibinfo {volume} {41}},\
  \bibinfo {pages} {164026} (\bibinfo {year} {2008})}\BibitemShut {NoStop}%
\bibitem [{\citenamefont {Sassaroli}\ \emph {et~al.}(1994)\citenamefont
  {Sassaroli}, \citenamefont {Srivastava},\ and\ \citenamefont
  {Widom}}]{sassaroli1994Photon}%
  \BibitemOpen
  \bibfield  {author} {\bibinfo {author} {\bibfnamefont {E.}~\bibnamefont
  {Sassaroli}}, \bibinfo {author} {\bibfnamefont {Y.~N.}\ \bibnamefont
  {Srivastava}}, \ and\ \bibinfo {author} {\bibfnamefont {A.}~\bibnamefont
  {Widom}},\ }\bibfield  {title} {\emph {\bibinfo {title} {Photon production by
  the dynamical casimir effect},\ }}\href {\doibase 10.1103/PhysRevA.50.1027}
  {\bibfield  {journal} {\bibinfo  {journal} {Phys. Rev. A}\ }\textbf {\bibinfo
  {volume} {50}},\ \bibinfo {pages} {1027} (\bibinfo {year}
  {1994})}\BibitemShut {NoStop}%
\bibitem [{\citenamefont {Haro}\ and\ \citenamefont
  {Elizalde}(2006)}]{haro2006Hamiltonian}%
  \BibitemOpen
  \bibfield  {author} {\bibinfo {author} {\bibfnamefont {J.}~\bibnamefont
  {Haro}}\ and\ \bibinfo {author} {\bibfnamefont {E.}~\bibnamefont
  {Elizalde}},\ }\bibfield  {title} {\emph {\bibinfo {title} {Hamiltonian
  approach to the dynamical casimir effect},\ }}\href {\doibase
  10.1103/PhysRevLett.97.130401} {\bibfield  {journal} {\bibinfo  {journal}
  {Phys. Rev. Lett.}\ }\textbf {\bibinfo {volume} {97}},\ \bibinfo {pages}
  {130401} (\bibinfo {year} {2006})}\BibitemShut {NoStop}%
\bibitem [{\citenamefont {Settineri}\ \emph {et~al.}(2019)\citenamefont
  {Settineri}, \citenamefont {Macr{\`i}}, \citenamefont {Garziano},
  \citenamefont {Di~Stefano}, \citenamefont {Nori},\ and\ \citenamefont
  {Savasta}}]{settineri2019Conversion}%
  \BibitemOpen
  \bibfield  {author} {\bibinfo {author} {\bibfnamefont {A.}~\bibnamefont
  {Settineri}}, \bibinfo {author} {\bibfnamefont {V.}~\bibnamefont
  {Macr{\`i}}}, \bibinfo {author} {\bibfnamefont {L.}~\bibnamefont {Garziano}},
  \bibinfo {author} {\bibfnamefont {O.}~\bibnamefont {Di~Stefano}}, \bibinfo
  {author} {\bibfnamefont {F.}~\bibnamefont {Nori}}, \ and\ \bibinfo {author}
  {\bibfnamefont {S.}~\bibnamefont {Savasta}},\ }\bibfield  {title} {\emph
  {\bibinfo {title} {Conversion of mechanical noise into correlated photon
  pairs: Dynamical casimir effect from an incoherent mechanical drive},\
  }}\href {\doibase 10.1103/PhysRevA.100.022501} {\bibfield  {journal}
  {\bibinfo  {journal} {Phys. Rev. A}\ }\textbf {\bibinfo {volume} {100}},\
  \bibinfo {pages} {022501} (\bibinfo {year} {2019})}\BibitemShut {NoStop}%
\bibitem [{\citenamefont {Lan}\ \emph {et~al.}(2024)\citenamefont {Lan},
  \citenamefont {Chen}, \citenamefont {Cheng}, \citenamefont {Chen},
  \citenamefont {Ye},\ and\ \citenamefont {Zhong}}]{lan2024Dynamical}%
  \BibitemOpen
  \bibfield  {author} {\bibinfo {author} {\bibfnamefont {Z.-L.}\ \bibnamefont
  {Lan}}, \bibinfo {author} {\bibfnamefont {Y.-W.}\ \bibnamefont {Chen}},
  \bibinfo {author} {\bibfnamefont {L.-Y.}\ \bibnamefont {Cheng}}, \bibinfo
  {author} {\bibfnamefont {L.}~\bibnamefont {Chen}}, \bibinfo {author}
  {\bibfnamefont {S.-Y.}\ \bibnamefont {Ye}}, \ and\ \bibinfo {author}
  {\bibfnamefont {Z.-R.}\ \bibnamefont {Zhong}},\ }\bibfield  {title} {\emph
  {\bibinfo {title} {Dynamical casimir effect in a hybrid cavity optomechanical
  system},\ }}\href {\doibase 10.1007/s11128-024-04267-3} {\bibfield  {journal}
  {\bibinfo  {journal} {Quantum Inf. Process.}\ }\textbf {\bibinfo {volume}
  {23}},\ \bibinfo {pages} {72} (\bibinfo {year} {2024})}\BibitemShut {NoStop}%
\bibitem [{\citenamefont {Wang}\ \emph {et~al.}(2023)\citenamefont {Wang},
  \citenamefont {Hu}, \citenamefont {Macr{\`i}}, \citenamefont {Xiang},\ and\
  \citenamefont {Nori}}]{wang2023Coherent}%
  \BibitemOpen
  \bibfield  {author} {\bibinfo {author} {\bibfnamefont {B.}~\bibnamefont
  {Wang}}, \bibinfo {author} {\bibfnamefont {J.-M.}\ \bibnamefont {Hu}},
  \bibinfo {author} {\bibfnamefont {V.}~\bibnamefont {Macr{\`i}}}, \bibinfo
  {author} {\bibfnamefont {Z.-L.}\ \bibnamefont {Xiang}}, \ and\ \bibinfo
  {author} {\bibfnamefont {F.}~\bibnamefont {Nori}},\ }\bibfield  {title}
  {\emph {\bibinfo {title} {Coherent resonant coupling between atoms and a
  mechanical oscillator mediated by cavity-vacuum fluctuations},\ }}\href
  {\doibase 10.1103/PhysRevResearch.5.013075} {\bibfield  {journal} {\bibinfo
  {journal} {Phys. Rev. Research}\ }\textbf {\bibinfo {volume} {5}},\ \bibinfo
  {pages} {013075} (\bibinfo {year} {2023})}\BibitemShut {NoStop}%
\bibitem [{\citenamefont {Garziano}\ \emph {et~al.}(2016)\citenamefont
  {Garziano}, \citenamefont {Macr{\`i}}, \citenamefont {Stassi}, \citenamefont
  {Di~Stefano}, \citenamefont {Nori},\ and\ \citenamefont
  {Savasta}}]{garziano2016One}%
  \BibitemOpen
  \bibfield  {author} {\bibinfo {author} {\bibfnamefont {L.}~\bibnamefont
  {Garziano}}, \bibinfo {author} {\bibfnamefont {V.}~\bibnamefont {Macr{\`i}}},
  \bibinfo {author} {\bibfnamefont {R.}~\bibnamefont {Stassi}}, \bibinfo
  {author} {\bibfnamefont {O.}~\bibnamefont {Di~Stefano}}, \bibinfo {author}
  {\bibfnamefont {F.}~\bibnamefont {Nori}}, \ and\ \bibinfo {author}
  {\bibfnamefont {S.}~\bibnamefont {Savasta}},\ }\bibfield  {title} {\emph
  {\bibinfo {title} {One photon can simultaneously excite two or more atoms},\
  }}\href {\doibase 10.1103/PhysRevLett.117.043601} {\bibfield  {journal}
  {\bibinfo  {journal} {Phys. Rev. Lett.}\ }\textbf {\bibinfo {volume} {117}},\
  \bibinfo {pages} {043601} (\bibinfo {year} {2016})}\BibitemShut {NoStop}%
\bibitem [{\citenamefont {Macr{\`i}}\ \emph
  {et~al.}(2018{\natexlab{b}})\citenamefont {Macr{\`i}}, \citenamefont {Nori},\
  and\ \citenamefont {Kockum}}]{macri2018Simple}%
  \BibitemOpen
  \bibfield  {author} {\bibinfo {author} {\bibfnamefont {V.}~\bibnamefont
  {Macr{\`i}}}, \bibinfo {author} {\bibfnamefont {F.}~\bibnamefont {Nori}}, \
  and\ \bibinfo {author} {\bibfnamefont {A.~F.}\ \bibnamefont {Kockum}},\
  }\bibfield  {title} {\emph {\bibinfo {title} {Simple preparation of bell and
  greenberger-horne-zeilinger states using ultrastrong-coupling circuit qed},\
  }}\href {\doibase 10.1103/PhysRevA.98.062327} {\bibfield  {journal} {\bibinfo
   {journal} {Phys. Rev. A}\ }\textbf {\bibinfo {volume} {98}},\ \bibinfo
  {pages} {062327} (\bibinfo {year} {2018}{\natexlab{b}})}\BibitemShut
  {NoStop}%
\bibitem [{\citenamefont {Qi}\ and\ \citenamefont
  {Jing}(2020)}]{qi2020Generating}%
  \BibitemOpen
  \bibfield  {author} {\bibinfo {author} {\bibfnamefont {S.-f.}\ \bibnamefont
  {Qi}}\ and\ \bibinfo {author} {\bibfnamefont {J.}~\bibnamefont {Jing}},\
  }\bibfield  {title} {\emph {\bibinfo {title} {Generating noon states in
  circuit qed using a multiphoton resonance in the presence of counter-rotating
  interactions},\ }}\href {\doibase 10.1103/PhysRevA.101.033809} {\bibfield
  {journal} {\bibinfo  {journal} {Phys. Rev. A}\ }\textbf {\bibinfo {volume}
  {101}},\ \bibinfo {pages} {033809} (\bibinfo {year} {2020})}\BibitemShut
  {NoStop}%
\bibitem [{\citenamefont {Gr{\"o}blacher}\ \emph {et~al.}(2009)\citenamefont
  {Gr{\"o}blacher}, \citenamefont {Hammerer}, \citenamefont {Vanner},\ and\
  \citenamefont {Aspelmeyer}}]{groblacher2009Observation}%
  \BibitemOpen
  \bibfield  {author} {\bibinfo {author} {\bibfnamefont {S.}~\bibnamefont
  {Gr{\"o}blacher}}, \bibinfo {author} {\bibfnamefont {K.}~\bibnamefont
  {Hammerer}}, \bibinfo {author} {\bibfnamefont {M.~R.}\ \bibnamefont
  {Vanner}}, \ and\ \bibinfo {author} {\bibfnamefont {M.}~\bibnamefont
  {Aspelmeyer}},\ }\bibfield  {title} {\emph {\bibinfo {title} {Observation of
  strong coupling between a micromechanical resonator and an optical cavity
  field},\ }}\href {\doibase 10.1038/nature08171} {\bibfield  {journal}
  {\bibinfo  {journal} {Nature (London)}\ }\textbf {\bibinfo {volume} {460}},\
  \bibinfo {pages} {724} (\bibinfo {year} {2009})}\BibitemShut {NoStop}%
\bibitem [{\citenamefont {Verhagen}\ \emph {et~al.}(2012)\citenamefont
  {Verhagen}, \citenamefont {Del{\'e}glise}, \citenamefont {Weis},
  \citenamefont {Schliesser},\ and\ \citenamefont
  {Kippenberg}}]{verhagen2012Quantumcoherent}%
  \BibitemOpen
  \bibfield  {author} {\bibinfo {author} {\bibfnamefont {E.}~\bibnamefont
  {Verhagen}}, \bibinfo {author} {\bibfnamefont {S.}~\bibnamefont
  {Del{\'e}glise}}, \bibinfo {author} {\bibfnamefont {S.}~\bibnamefont {Weis}},
  \bibinfo {author} {\bibfnamefont {A.}~\bibnamefont {Schliesser}}, \ and\
  \bibinfo {author} {\bibfnamefont {T.~J.}\ \bibnamefont {Kippenberg}},\
  }\bibfield  {title} {\emph {\bibinfo {title} {Quantum-coherent coupling of a
  mechanical oscillator to an optical cavity mode},\ }}\href {\doibase
  10.1038/nature10787} {\bibfield  {journal} {\bibinfo  {journal} {Nature
  (London)}\ }\textbf {\bibinfo {volume} {482}},\ \bibinfo {pages} {63}
  (\bibinfo {year} {2012})}\BibitemShut {NoStop}%
\bibitem [{\citenamefont {Bochmann}\ \emph {et~al.}(2013)\citenamefont
  {Bochmann}, \citenamefont {Vainsencher}, \citenamefont {Awschalom},\ and\
  \citenamefont {Cleland}}]{bochmann2013Nanomechanical}%
  \BibitemOpen
  \bibfield  {author} {\bibinfo {author} {\bibfnamefont {J.}~\bibnamefont
  {Bochmann}}, \bibinfo {author} {\bibfnamefont {A.}~\bibnamefont
  {Vainsencher}}, \bibinfo {author} {\bibfnamefont {D.~D.}\ \bibnamefont
  {Awschalom}}, \ and\ \bibinfo {author} {\bibfnamefont {A.~N.}\ \bibnamefont
  {Cleland}},\ }\bibfield  {title} {\emph {\bibinfo {title} {Nanomechanical
  coupling between microwave and optical photons},\ }}\href {\doibase
  10.1038/nphys2748} {\bibfield  {journal} {\bibinfo  {journal} {Nat. Phys.}\
  }\textbf {\bibinfo {volume} {9}},\ \bibinfo {pages} {712} (\bibinfo {year}
  {2013})}\BibitemShut {NoStop}%
\bibitem [{\citenamefont {Aspelmeyer}\ \emph {et~al.}(2014)\citenamefont
  {Aspelmeyer}, \citenamefont {Kippenberg},\ and\ \citenamefont
  {Marquardt}}]{aspelmeyer2014Cavity}%
  \BibitemOpen
  \bibfield  {author} {\bibinfo {author} {\bibfnamefont {M.}~\bibnamefont
  {Aspelmeyer}}, \bibinfo {author} {\bibfnamefont {T.~J.}\ \bibnamefont
  {Kippenberg}}, \ and\ \bibinfo {author} {\bibfnamefont {F.}~\bibnamefont
  {Marquardt}},\ }\bibfield  {title} {\emph {\bibinfo {title} {Cavity
  optomechanics},\ }}\href {\doibase 10.1103/RevModPhys.86.1391} {\bibfield
  {journal} {\bibinfo  {journal} {Rev. Mod. Phys.}\ }\textbf {\bibinfo {volume}
  {86}},\ \bibinfo {pages} {1391} (\bibinfo {year} {2014})}\BibitemShut
  {NoStop}%
\bibitem [{\citenamefont {Korzh}\ \emph {et~al.}(2020)\citenamefont {Korzh},
  \citenamefont {Zhao}, \citenamefont {Allmaras}, \citenamefont {Frasca},
  \citenamefont {Autry}, \citenamefont {Bersin}, \citenamefont {Beyer},
  \citenamefont {Briggs}, \citenamefont {Bumble}, \citenamefont {Colangelo},
  \citenamefont {Crouch}, \citenamefont {Dane}, \citenamefont {Gerrits},
  \citenamefont {Lita}, \citenamefont {Marsili}, \citenamefont {Moody},
  \citenamefont {Pe{\~n}a}, \citenamefont {Ramirez}, \citenamefont {Rezac},
  \citenamefont {Sinclair}, \citenamefont {Stevens}, \citenamefont {Velasco},
  \citenamefont {Verma}, \citenamefont {Wollman}, \citenamefont {Xie},
  \citenamefont {Zhu}, \citenamefont {Hale}, \citenamefont {Spiropulu},
  \citenamefont {Silverman}, \citenamefont {Mirin}, \citenamefont {Nam},
  \citenamefont {Kozorezov}, \citenamefont {Shaw},\ and\ \citenamefont
  {Berggren}}]{korzh2020Demonstration}%
  \BibitemOpen
  \bibfield  {author} {\bibinfo {author} {\bibfnamefont {B.}~\bibnamefont
  {Korzh}}, \bibinfo {author} {\bibfnamefont {Q.-Y.}\ \bibnamefont {Zhao}},
  \bibinfo {author} {\bibfnamefont {J.~P.}\ \bibnamefont {Allmaras}}, \bibinfo
  {author} {\bibfnamefont {S.}~\bibnamefont {Frasca}}, \bibinfo {author}
  {\bibfnamefont {T.~M.}\ \bibnamefont {Autry}}, \bibinfo {author}
  {\bibfnamefont {E.~A.}\ \bibnamefont {Bersin}}, \bibinfo {author}
  {\bibfnamefont {A.~D.}\ \bibnamefont {Beyer}}, \bibinfo {author}
  {\bibfnamefont {R.~M.}\ \bibnamefont {Briggs}}, \bibinfo {author}
  {\bibfnamefont {B.}~\bibnamefont {Bumble}}, \bibinfo {author} {\bibfnamefont
  {M.}~\bibnamefont {Colangelo}}, \bibinfo {author} {\bibfnamefont {G.~M.}\
  \bibnamefont {Crouch}}, \bibinfo {author} {\bibfnamefont {A.~E.}\
  \bibnamefont {Dane}}, \bibinfo {author} {\bibfnamefont {T.}~\bibnamefont
  {Gerrits}}, \bibinfo {author} {\bibfnamefont {A.~E.}\ \bibnamefont {Lita}},
  \bibinfo {author} {\bibfnamefont {F.}~\bibnamefont {Marsili}}, \bibinfo
  {author} {\bibfnamefont {G.}~\bibnamefont {Moody}}, \bibinfo {author}
  {\bibfnamefont {C.}~\bibnamefont {Pe{\~n}a}}, \bibinfo {author}
  {\bibfnamefont {E.}~\bibnamefont {Ramirez}}, \bibinfo {author} {\bibfnamefont
  {J.~D.}\ \bibnamefont {Rezac}}, \bibinfo {author} {\bibfnamefont
  {N.}~\bibnamefont {Sinclair}}, \bibinfo {author} {\bibfnamefont {M.~J.}\
  \bibnamefont {Stevens}}, \bibinfo {author} {\bibfnamefont {A.~E.}\
  \bibnamefont {Velasco}}, \bibinfo {author} {\bibfnamefont {V.~B.}\
  \bibnamefont {Verma}}, \bibinfo {author} {\bibfnamefont {E.~E.}\ \bibnamefont
  {Wollman}}, \bibinfo {author} {\bibfnamefont {S.}~\bibnamefont {Xie}},
  \bibinfo {author} {\bibfnamefont {D.}~\bibnamefont {Zhu}}, \bibinfo {author}
  {\bibfnamefont {P.~D.}\ \bibnamefont {Hale}}, \bibinfo {author}
  {\bibfnamefont {M.}~\bibnamefont {Spiropulu}}, \bibinfo {author}
  {\bibfnamefont {K.~L.}\ \bibnamefont {Silverman}}, \bibinfo {author}
  {\bibfnamefont {R.~P.}\ \bibnamefont {Mirin}}, \bibinfo {author}
  {\bibfnamefont {S.~W.}\ \bibnamefont {Nam}}, \bibinfo {author} {\bibfnamefont
  {A.~G.}\ \bibnamefont {Kozorezov}}, \bibinfo {author} {\bibfnamefont {M.~D.}\
  \bibnamefont {Shaw}}, \ and\ \bibinfo {author} {\bibfnamefont {K.~K.}\
  \bibnamefont {Berggren}},\ }\bibfield  {title} {\emph {\bibinfo {title}
  {Demonstration of sub-3 ps temporal resolution with a superconducting
  nanowire single-photon detector},\ }}\href {\doibase
  10.1038/s41566-020-0589-x} {\bibfield  {journal} {\bibinfo  {journal} {Nat.
  Photonics}\ }\textbf {\bibinfo {volume} {14}},\ \bibinfo {pages} {250}
  (\bibinfo {year} {2020})}\BibitemShut {NoStop}%
\bibitem [{\citenamefont {Rettaroli}\ \emph {et~al.}(2025)\citenamefont
  {Rettaroli}, \citenamefont {Banchi}, \citenamefont {Corti}, \citenamefont
  {D'Elia}, \citenamefont {Gatti}, \citenamefont {Giachero}, \citenamefont
  {Labranca}, \citenamefont {Moretti}, \citenamefont {Nucciotti}, \citenamefont
  {Komnang},\ and\ \citenamefont {Tocci}}]{rettaroli2025Novel}%
  \BibitemOpen
  \bibfield  {author} {\bibinfo {author} {\bibfnamefont {A.}~\bibnamefont
  {Rettaroli}}, \bibinfo {author} {\bibfnamefont {L.}~\bibnamefont {Banchi}},
  \bibinfo {author} {\bibfnamefont {H.~A.}\ \bibnamefont {Corti}}, \bibinfo
  {author} {\bibfnamefont {A.}~\bibnamefont {D'Elia}}, \bibinfo {author}
  {\bibfnamefont {C.}~\bibnamefont {Gatti}}, \bibinfo {author} {\bibfnamefont
  {A.}~\bibnamefont {Giachero}}, \bibinfo {author} {\bibfnamefont
  {D.}~\bibnamefont {Labranca}}, \bibinfo {author} {\bibfnamefont
  {R.}~\bibnamefont {Moretti}}, \bibinfo {author} {\bibfnamefont
  {A.}~\bibnamefont {Nucciotti}}, \bibinfo {author} {\bibfnamefont {A.~S.~P.}\
  \bibnamefont {Komnang}}, \ and\ \bibinfo {author} {\bibfnamefont
  {S.}~\bibnamefont {Tocci}},\ }\bibfield  {title} {\emph {\bibinfo {title}
  {Novel two-qubit microwave photon detector for fundamental physics
  applications},\ }}\href {\doibase 10.1016/j.nima.2024.170010} {\bibfield
  {journal} {\bibinfo  {journal} {Nucl. Instrum. Methods Phys. Res., Sect. A}\
  }\textbf {\bibinfo {volume} {1070}},\ \bibinfo {pages} {170010} (\bibinfo
  {year} {2025})}\BibitemShut {NoStop}%
\bibitem [{\citenamefont {Pirkkalainen}\ \emph {et~al.}(2015)\citenamefont
  {Pirkkalainen}, \citenamefont {Cho}, \citenamefont {Massel}, \citenamefont
  {Tuorila}, \citenamefont {Heikkil{\"a}}, \citenamefont {Hakonen},\ and\
  \citenamefont {Sillanp{\"a}{\"a}}}]{pirkkalainen2015Cavity}%
  \BibitemOpen
  \bibfield  {author} {\bibinfo {author} {\bibfnamefont {J.-M.}\ \bibnamefont
  {Pirkkalainen}}, \bibinfo {author} {\bibfnamefont {S.}~\bibnamefont {Cho}},
  \bibinfo {author} {\bibfnamefont {F.}~\bibnamefont {Massel}}, \bibinfo
  {author} {\bibfnamefont {J.}~\bibnamefont {Tuorila}}, \bibinfo {author}
  {\bibfnamefont {T.}~\bibnamefont {Heikkil{\"a}}}, \bibinfo {author}
  {\bibfnamefont {P.}~\bibnamefont {Hakonen}}, \ and\ \bibinfo {author}
  {\bibfnamefont {M.}~\bibnamefont {Sillanp{\"a}{\"a}}},\ }\bibfield  {title}
  {\emph {\bibinfo {title} {Cavity optomechanics mediated by a quantum
  two-level system},\ }}\href {\doibase 10.1038/ncomms7981} {\bibfield
  {journal} {\bibinfo  {journal} {Nat. Commun.}\ }\textbf {\bibinfo {volume}
  {6}},\ \bibinfo {pages} {6981} (\bibinfo {year} {2015})}\BibitemShut
  {NoStop}%
\end{thebibliography}%
	
\end{document}